\documentclass[useAMS,usenatbib]{mn2e}
\usepackage{amsmath, amssymb, graphicx,times,rotating}
\usepackage{natbib}

\usepackage{float}
\usepackage{placeins}

\title{Frequency dependent core shifts and parameter estimation for the blazar 3C 454.3}

\author[Mohan et al.]
{P. Mohan$^{1}$\thanks{E-mail: pmohan@aries.res.in}, 
A. Agarwal$^{1,2}$,
A. Mangalam$^{3}$,
Alok C.\ Gupta$^{1,2,4}$\thanks{CAS Visiting Fellow},
Paul J. Wiita$^{5}$,
A.\ E.\ Volvach$^{6,7}$,
\newauthor M.\ F.\ Aller$^{8}$,
H.\ D.\ Aller$^{8}$,
M. F. Gu$^{4}$,
A.\ L{\"a}hteenm{\"a}ki$^{9}$,
M.\ Tornikoski$^{9}$, 
L.\ N.\ Volvach$^{6,7}$
\\ \\
$^{1}$Aryabhatta Research Institute of Observational Sciences (ARIES),
Manora Peak, Nainital - 263002, India.\\
$^{2}$Department of Physics, DDU Gorakhpur University, Gorakhpur - 273009, India. \\
$^{3}$Indian Institute of Astrophysics, Sarjapur Road, Koramangala, Bangalore - 560034, India. \\
$^{4}$Key Laboratory for Research in Galaxies and Cosmology, Shanghai Astronomical Observatory, Chinese Academy of Sciences, Shanghai 200030.\\
$^{5}$Department of Physics, The College of New Jersey, P.O.\ Box 7718, Ewing, NJ 08628, U.S.A.\\
$^{6}$Radio Astronomy Laboratory of Crimean Astrophysical Observatory, Crimea. \\
$^{7}$Taras Shevchenko National University of Kyiv, Kiev, Ukraine. \\
$^{8}$Astronomy Department, University of Michigan, Ann Arbor, MI, U.S.A.\\
$^{9}$Aalto University Mets{\"a}hovi Radio Observatory, Finland.\\ 
}

\begin{document}

\date{Accepted 2015 June 22; Received 2015 June 11; in original form 2015 March 12.}

\pubyear{2015}

\maketitle

\label{firstpage}

\begin{abstract}

We study the core shift effect in the parsec scale jet of the blazar 3C 454.3 using the 4.8 GHz - 36.8 GHz radio light curves obtained from three decades of continuous monitoring. From a piecewise Gaussian fit to each flare, time lags $\Delta t$ between the observation frequencies $\nu$ and spectral indices $\alpha$ based on peak amplitudes $A$ are determined. From the fit $\Delta t \propto \nu^{1/k_r}$, {\bf $k_r = 1.10 \pm 0.18$} indicating equipartition between the magnetic field energy density and the particle energy density. From the fit $A \propto \nu^\alpha$, $\alpha$ is in the range $-0.24$ to $1.52$. A mean magnetic field strength at 1 pc, $B_1 = 0.5 \pm 0.2$ G, and at the core, $B_{\rm core} = 46 \pm 16$ mG, are inferred, consistent with previous estimates. The measure of core position offset is $\Omega_{r\nu} = 6.4 \pm 2.8$ pc GHz$^{1/k_r}$ when averaged over all frequency pairs. Based on the statistical trend shown by the measured core radius $r_{\rm core}$ as a function of $\nu$, we infer that the synchrotron opacity model may not be valid for all cases. A Fourier periodogram analysis yields power law slopes in the range $-1.6$ to $-3.5$ describing the power spectral density shape and gives bend timescales in the range $0.52 - 0.66~$yr. This result, and both positive and negative $\alpha$, indicate that the flares originate from multiple shocks in a small region. Important objectives met in our study include: the demonstration of the computational efficiency and statistical basis of the piecewise Gaussian fit; consistency with previously reported results; evidence for the core shift dependence on observation frequency and its utility in jet diagnostics in the region close to the resolving limit of very long baseline interferometry observations.
\end{abstract}


\section{Introduction} \label{sec:intro}

Two groups of objects, BL Lacertae (BL Lac; with featureless optical spectra) and flat spectrum radio quasar (FSRQ; having prominent emission lines) constitute a subset of radio-loud Active Galactic Nuclei (RL-AGNs) called blazars, which according to the orientation based unified
schemes of RL AGNs have jets making an angle of $\leq$ 10$^{\circ}$ or so to the line of sight (e.g. \citealt{1995PASP..107..803U}). AGNs are characterized by strong flux variability at all wavelengths, strong and variable polarization in the radio to optical wavelengths, and a core dominated radio spectrum. Blazar emission from the relativistic jet is dominated by non thermal emission (synchrotron and inverse Compton) which is Doppler boosted making a cone angle of $\sim 1/\Gamma$ where $\Gamma$ is the bulk Lorentz factor. Flicker in radio flux from these AGN can be caused by interstellar scintillation. The apparent extremely high brightness temperatures of $> 10^{17}$K inferred from radio observations can be attributed to relativistic beaming, coherent radiation mechanisms and geometric effects (\citealt{1995ARA&A..33..163W}). Periodic variability in the blazar might be caused by the helical movement of the jet components, shocks, jet precession effects, accretion disc instabilities (\citealt{1991PASJ...43..261H,1992A&A...255...59C, 1992A&A...259..109G,1997ApJ...482L..33G,2006A&A...452...83B,2010A&A...511A..57B}) or due to the rotation of the secondary black hole around the primary supermassive black hole (e.g. \citealt{1988ApJ...325..628S,2008Natur.452..851V,2000A&A...353..473R}). Doppler boosting shortens the variability timescales by a factor of $\delta^{-1}$ where this Doppler factor depends on the Lorentz factor as $\delta = 1/\Gamma(1-\beta \cos \theta)$ where $\theta$ is the angle between the jet direction and the observer line of sight and $\beta = (1-\Gamma^{-2})^{1/2}$ is the speed of the jet component in the source frame.

3C 454.3 (PKS 2251+158) is a highly variable blazar, classified as a FSRQ with redshift $z = 0.859$ (e.g. \citealt{1989QSO...C......0H,2001ApJS..134..181J}) with a black hole mass of $4.4 \times 10^9 ~ M_\odot$ (\citealt{2001MNRAS.327.1111G}). An ultra-violet excess due to the dominance of thermal emission from the accretion disc over synchrotron radiation is inferred in \cite{2007A&A...473..819R,2008A&A...480..339R}. Since 2001, the FSRQ has displayed strong variability over the entire electromagnetic spectrum. It underwent a bright outburst almost simultaneously from radio to X-ray bands in May 2005 when it reached its brightest R band magnitude of 12.0 (\citealt{2006A&A...445L...1F,2010ATel.2995....1V,2006A&A...453..817V,2007A&A...464L...5V,2008A&A...480..339R,2010ApJ...715..362J}). Flux variations at 230 GHz were detected around two months after the optical variations. This outburst was inferred to be caused by shocks propagating along the relativistic jet when the emitting region at optical wavelength passes closest to the observer's line of sight first and does so later on for other wavelengths (\citealt{2007A&A...464L...5V}). The 2005 outburst was followed by a comparatively quiescent period till spring 2007 at all frequencies, when it displayed a ``big blue bump'', characteristic of radio-quiet AGNs and indicating thermal emission from the accretion disc (\citealt{2007A&A...473..819R}). \cite{2007A&A...463..529K} proposed a model to explain the activities of the FSRQ during 2005-2007 time period, according to which the location from which bulk kinetic energy is dissipated along the jet is dependent on the bulk Lorentz factor and compactness of the perturbations which arise. Correlations between light curves at different frequencies during the 2005-2006 flare established that a single radiation mechanism is responsible for both radio and optical emission. Observations were carried out on all possible timescales, i.e., from days to years, in both optical and radio from which it was found that the duration of the flare and the individual features of the flare were same in both, thus showing that both were produced by the Doppler boosted relativistic jet.

A multiwavelength monitoring campaign during a high flux state in 2008 (\citealt{2009ApJ...697L..81B}) indicates a strong correlation between infrared, optical, ultraviolet and $\gamma$-ray wavelengths. This is attributed to the external Compton scattering where accretion disk and emission line photons are scattered by strongly relativistic electrons (with Lorentz factors $10^3 - 10^4$) which are also radiating synchrotron based infrared and optical emission. A study of intra-day variability in the source during 2009 - 2010 (\citealt{2012AJ....143...23G}) during a high flux state indicates a strong correlation between the $\gamma$-ray and optical wavelengths and that the former leads the latter by $4.5 \pm 1.0$ days and is attributed to external Compton scattering. A multi frequency study from millimeter wavelengths to $\gamma$-rays during three outburst events in 2009 - 2010 (\citealt{2013ApJ...773..147J}) reveals a similar radiation mechanism and location of the flares with small to no time lag being found between optical and $\gamma$-ray light curves. Typical variability timescales are inferred to be $\sim$ 3 hours in the parsec scale jet adding to the conclusion that the emission is highly localized. The measured magnetic field strength is in the range $0.1 - 1$ G. A study to derive the bulk kinetic power using jet parameters derived in the framework of the \cite{1981ApJ...243..700K} inhomogeneous jet model at the 43 GHz observational frequency indicates a jet viewing angle of 3.8$^{\circ}$, bulk Lorentz factor of 32.4 and Doppler factor of 20.3 (\citealt{2009MNRAS.396..984G}). From the derived bulk kinetic luminosity and the luminosity from the broad line emission (due to disk based sources), the study infers a strong correlation between these two indicating that the jet power is intrinsically linked to the accretion process. Owing to its high flux state and broadband spectral energy distribution (SED), it offers opportunities for intra-day variability (IDV) studies (\citealt{2010MNRAS.405L..94T}) as well as spectral variability studies. A study of the relationship between the bulk Lorentz factor and the AGN properties (\citealt{2012ApJ...759..114C}) estimates jet parameters derived in the framework of the inhomogeneous jet model \citep{1981ApJ...243..700K}. A magnetic field strength at 1 pc of 0.493 G was obtained, and from the large sample of radio-loud AGN, a strong correlation is inferred between the bulk Lorentz factor and the black hole mass, indicating that accretion and the resulting black hole spin up could play a role in jet power.

In section \ref{coreshifts}, we describe parameters that can be derived from the analysis of the radio light curves, including the magnetic field strength and size of emitting core, based on physical models proposed in earlier literature. In section \ref{observations}, we describe the observations made to obtain the final analyzable light curves and the analysis procedure which includes a piecewise Gaussian fit to each flaring portion in the light curve. In this section, we then describe the time series analysis of each light curve segment using the Fourier periodogram. In section \ref{results}, the results of the analysis and the estimated parameters including the time lags, spectral indices, magnetic field strength, size of emitting core and others are presented. This is followed by a discussion of the results from the time series analysis. We finally conclude in section \ref{conclusions} with the comparison of our results with earlier work and its impact.

\section{Frequency-dependent core shifts}
\label{coreshifts}

Due to finite VLBI resolution, the true optically thick radio core (i.e. surface with optical depth $\tau = 1$) is not completely resolved.
Since the $\tau=1$ surface is frequency dependent (\citealt{1981ApJ...243..700K,2009MNRAS.400...26O}), the radio core position depends on the observation frequency $\nu$ as $r\propto \nu^{-1/k_{r}}$ (\citealt{1981ApJ...243..700K}), with $r$ being the distance from the central region and the index $k_{r}$ given by (\citealt{1998A&A...330...79L}),
\begin{equation}
k_{r} = \frac{(3 - 2\alpha)m + 2n - 2}{5 - 2\alpha}
\label{k_r}
\end{equation}
under the assumption that the ambient medium pressure on the jet is negligible (\citealt{1981ApJ...243..700K,1998A&A...330...79L}) and external pressure is non-negligible with non-zero gradient along the jet. In the above expression for $k_{r}$, $\alpha$ is the spectral index ($S\propto \nu^{\alpha}$) while $m$ and $n$ describe the power law decay of the magnetic field ($B = B_{1} (r/1 \rm{pc})^{-m}$) and particle number density ($N = N_{1} (r/1 \rm{pc})^{-n}$) respectively, with distance from the jet base (i.e.  the jet apex). Here $N_{1}$ and $B_{1}$ refer to values of $N$ and $B$ at a distance of $r = 1$ pc from the core in the observer's frame. These power law expressions are valid till $\sim$ 1 pc as the jet shape changes from conical to parabolic closer to the true jet origin (e.g., \citealt{2006evn..confE...2K}; \citealt{2012ApJ...745L..28A}; \citealt{2013ApJ...775..118N}; \citealt{2006MNRAS.367..375B}; \citealt{2006MNRAS.368.1561M}). Assuming equipartition between electron energy density and the magnetic energy density we obtain $n = 2m$  (\citealt{1998A&A...330...79L}). Assuming the combination $m = 1$ and $n = 2$ which describes well the synchrotron emission from compact VLBI cores (\citealt{1981ApJ...243..700K,1998A&A...330...79L}) for constant jet flow speed along with constant opening angle, we obtain $k_r = 1$, independent of the spectral index $\alpha$.

In the shock-in-jet model scenario, shocks are produced as a result of change in physical quantities such as velocity, electron density, magnetic field through the jet or pressure at the base of the jet. These shocks propagate down the jet and cause core brightening, thus resulting in a flaring state of the source which is followed by the appearance of a jet component in VLBI images (e.g. \citealt{1985ApJ...298..114M,1997ApJ...482L..33G}). Let R$_{on}$ be the distance of the onset of a shock which crosses the $\tau=1$ surface at time $T_i$ for frequency $\nu_{i}$ thus giving a flare in the light curve. The core position shift with frequency gives time lags between different frequency light curves. For apparent superluminal motion (e.g. \citealt{1967MNRAS.135..345R,2000A&A...361..850T}), the time $t$ in the observer's frame passed after the onset of the shock wave at the distance $R_{on}$ is (\citealt{2011MNRAS.415.1631K})
\begin{equation}
t = \frac{(1 + z) \sin \theta(R(\nu) - R_{on}(\nu))}{\beta_{app} c},
\end{equation}
where $R$ is the distance along the jet axis, $\beta_{app}$ is the apparent superluminal velocity of the jet component and $\theta$ is the jet viewing angle.

The time lag between light curves at two frequencies is given by
\begin{align}
\Delta t & = t(\nu_{a}) - t(\nu_{b})\\ \nonumber
   & =\frac{(1+z) \sin \theta}{\beta_{app} c} [R(\nu_{a}) - R(\nu_{b}) +
R_{on}(\nu_{a}) - R_{on}(\nu_{b})].
\end{align}
Since $r\propto$ $\nu^{-1/k_{r}}$, $\Delta t \propto \nu^{-1/k_{r}}$. Hence, the determination of time lags from the light curves at various frequencies can be used to determine the index $k_r$.

$\Omega_{r\nu}$ is a measure of the core-position offset in pc(GHz)$^{1/k_r}$ given by (\citealt{1998A&A...330...79L})
\begin{equation}
\Omega_{r\nu}=4.85\times10^{-9} \frac{\Delta r_{\rm mas} D_L}{(1+z)^2} \left( \frac{\nu_1^{1/k_r}\nu_2^{1/k_r}}{\nu_2^{1/k_r}-\nu_1^{1/k_r}} \right),
\label{omega}
\end{equation}
where $\Delta r_{\rm mas} = \mu \Delta t$ is the core-position offset (in milli-arcseconds) between two frequencies $\nu_1$ and $\nu_2$ (in GHz) in terms of the proper motion $\mu$ and D$_L$ is the luminosity distance in pc. The distance of VLBI core from the base of the jet at frequency $\nu$ is given by (\citealt{1998A&A...330...79L})
\begin{equation}
r_{core}(\nu) = \frac{\Omega_{r\nu}}{\sin \theta} \nu^{-1/k_{r}}.
 \label{rcore}
\end{equation}
Using Eqn.\ (43) from \cite{2005ApJ...619...73H} along with the assumption of equipartition between magnetic field energy density and particle energy density, the magnetic field strength in Gauss at 1 pc from the jet base is given by
\begin{equation} 
B_{1} \cong 0.025 \left(\frac{\Omega_{r\nu}^3 (1+z)^2}{\delta^2 \varphi \sin^2 \theta}\right)^{1/4},
\label{B1}
\end{equation}
where $\theta$ is the jet viewing angle, $\varphi$ is the jet half-opening angle, $\delta$ is the Doppler factor, $z$ is the cosmological redshift and $\Omega_{r\nu}$ is the core-position offset defined in Eqn.\ (\ref{omega}). Using the above equations and the assumption that $m = 1$ for an equipartition magnetic field,
\begin{equation} 
B_{core}(\nu) = B_{1} r_{core}^{-1}.
\label{Bcore}
\end{equation}

\section{Observations and analysis technique}
\label{observations}

The long term light curves of 3C 454.3 at 4.8 GHz, 8.0 GHz and 14.5 GHz are obtained from the University of Michigan Radio Astronomical Observatory (UMRAO) (\citealt{1999ASPC..159...45A}). The UMRAO data at 4.8 GHz spans the period between 1978 and 2013, while the 8.0 GHz data spans from 1966 to 2013 and the 14.5 GHz data covers more than 35 years from 1973 to 2013. Radio monitoring at 22.2 and 36.8 GHz were carried out with the 22-m radio telescope (RT-22) of the Crimean Astrophysical Observatory (CrAO) (\citealt{2006ASPC..360..133V}). We use modulated radiometers in combination with the registration regime ``ON-ON" for collecting data from the telescope (\citealt{2000AstL...26..204N}). Observations at 37.0 GHz were made with the 14 m radio telescope of Aalto University Mets\"ahovi Radio Observatory in Finland. Data obtained at Mets\"ahovi and RT-22 were combined in a single array to supplement each other. A detailed description of the data reduction and analysis of Mets\"ahovi data is given in \cite{1998A&AS..132..305T}.

\begin{figure}
\centering
\includegraphics[scale=0.48]{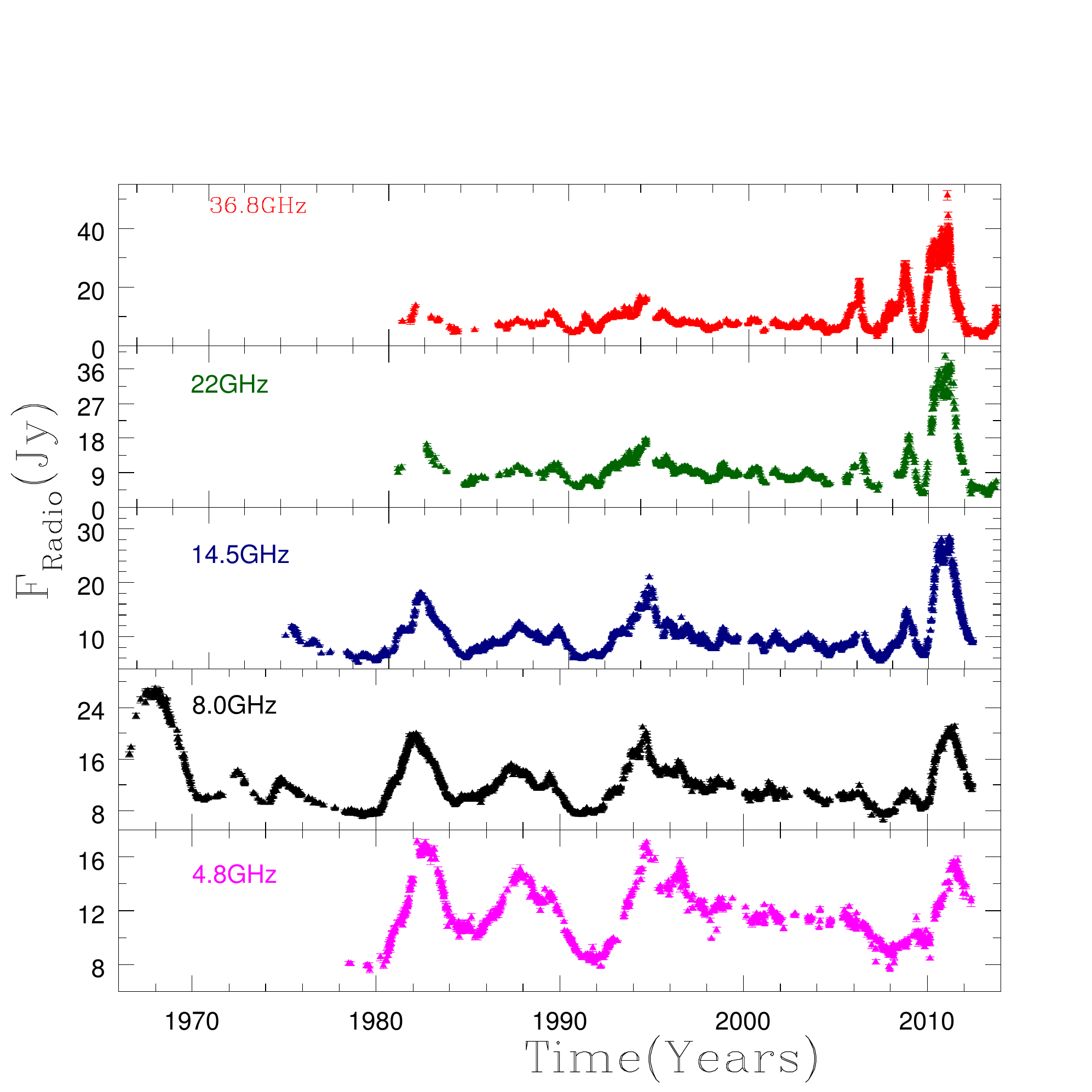}
\caption{Long-term variability light curves of 3C 454.3 in the 4.8 GHz - 36.8 GHz frequency range.}
\label{ltvfigure}
\end{figure}

The light curves of 3C 454.3 from 4.8 GHz to 36.8 GHz are displayed in Fig. \ref{ltvfigure}. Long term monitoring is necessary for studying and analyzing various properties of AGNs such as the dynamical evolution, the inner sub-parsec structure, radiation mechanisms and location of radiating regions.

The light curve is subjected to pre-processing in order to determine the positions of the maxima and minima which represent the boundaries of the flaring portions. A Gaussian filtering with a suitable parameter is applied to filter high frequency components and hence determine the extrema in the light curve. Then, a Gaussian function of the form
\begin{equation}
y = A e^{[-(t-\overline{m})^2/(2 \sigma^2)]},
\label{Gaussian}
\end{equation}
is used to represent the shape of each flare region in the light curve. The fit is carried out from a base line at the minimum amplitude of the
light curve. Initial values $A_0$, $\overline{m}_0$ and $\sigma_0$ are taken to be the maxima values, maxima positions and half the differences
between consecutive minima (assuming that a flare would be symmetric) respectively. As $A_0$, $\overline{m}_0$ and $\sigma_0$ are for a smoothed version
of the raw light curve, trial values for $A$, $\overline{m}$ and $\sigma$ are generated based on regions around these initial values. These trial values
account for possible values in a region around the flare peak amplitude, positions and the differences in the original light curve. A matrix of $(A_i,\overline{m}_i,\sigma_i)$ is constructed for all combinations of trial values of each peak. Each three parameter combination is fed into
eqn. (\ref{Gaussian}) along with the appropriate $t$ spanning each flare. This procedure is cycled through the entire light curve such that all
flares are covered by the Gaussian fit. For each Gaussian section covering a flare
\begin{equation}
\chi^2_i = \sum_i \frac{(x_i (t_i)-y(A_i,\overline{m}_i,\sigma_i))^2}{{\rm Variance}(x_i (t_i))},
\end{equation}
is determined for each section between the minima and the best fit for that section would be that combination $(A_i,\overline{m}_i,\sigma_i)$ which gives the least value of $\chi^2_i$. The $1~\sigma$ errors are determined for each parameter based on the table of $\chi^2_i$ generated as the distance from the best fit minimum $\chi^2$ value. 

\begin{figure}
\centerline{\includegraphics[scale=0.33]{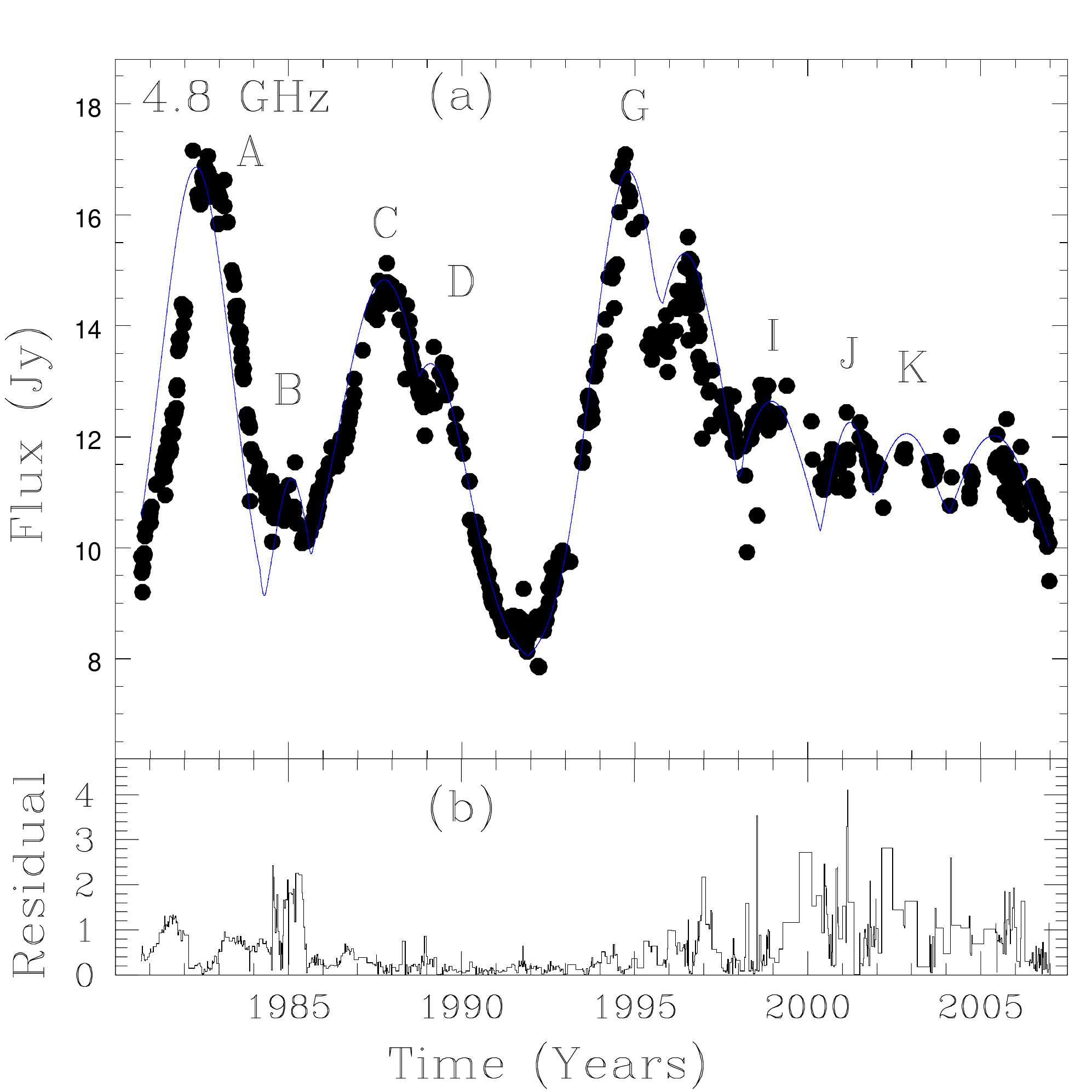}}
\centerline{\includegraphics[scale=0.33]{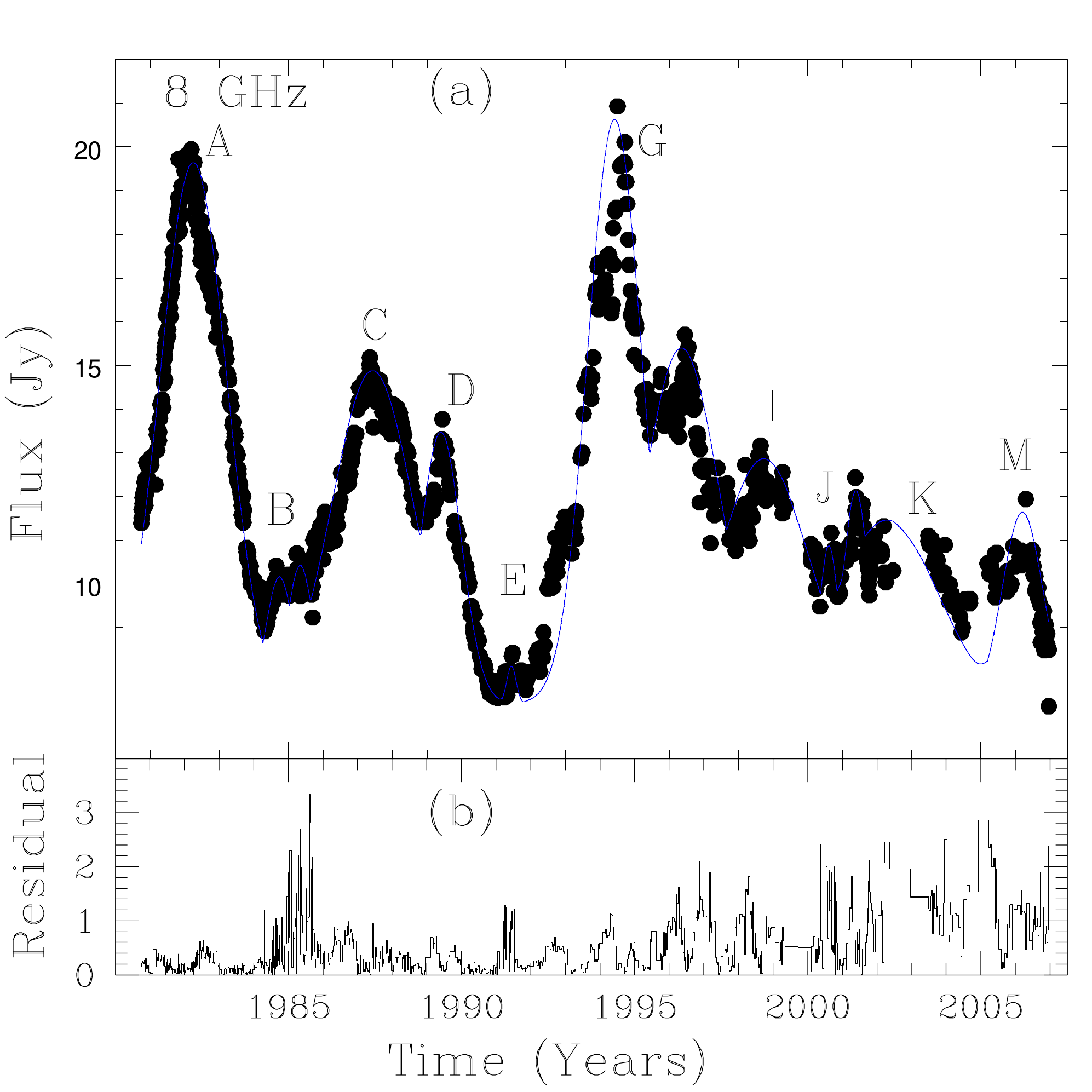}}
\centerline{\includegraphics[scale=0.33]{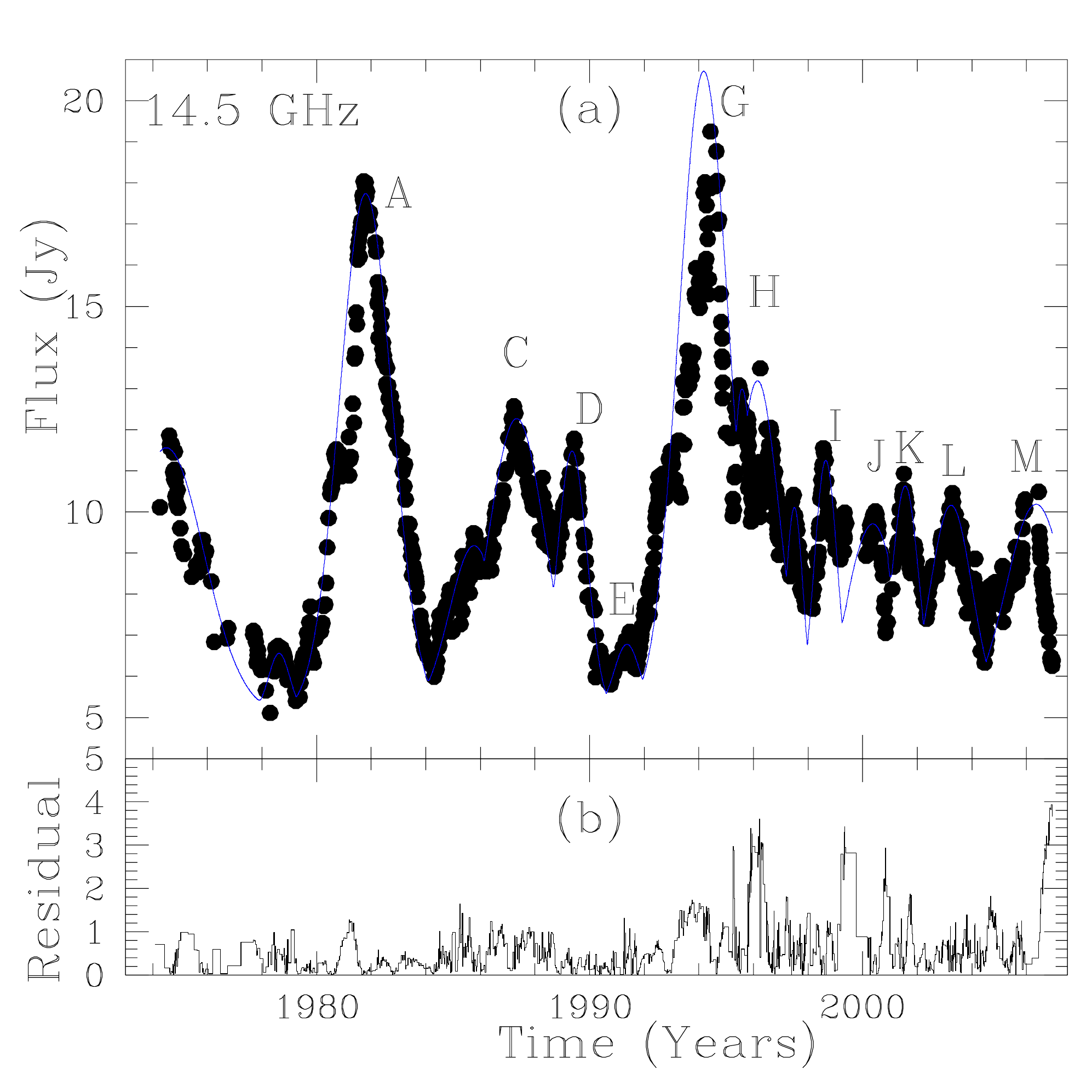}}
\caption{Segment 1 (beginning of observations to $\sim$ 2007.0) of the 4.8 GHz, 8.0 GHz and 14.5 GHz light curve flares fit with a piecewise Gaussian function. The residual in the lower panels are calculated as $[y(A_i,\overline{m}_i,\sigma_i)-x_i(t_i)]/{\rm Standard ~Deviation(x_i (t_i))}$.}
\label{seg1ga}
\end{figure}

\begin{figure}
\centerline{\includegraphics[scale=0.33]{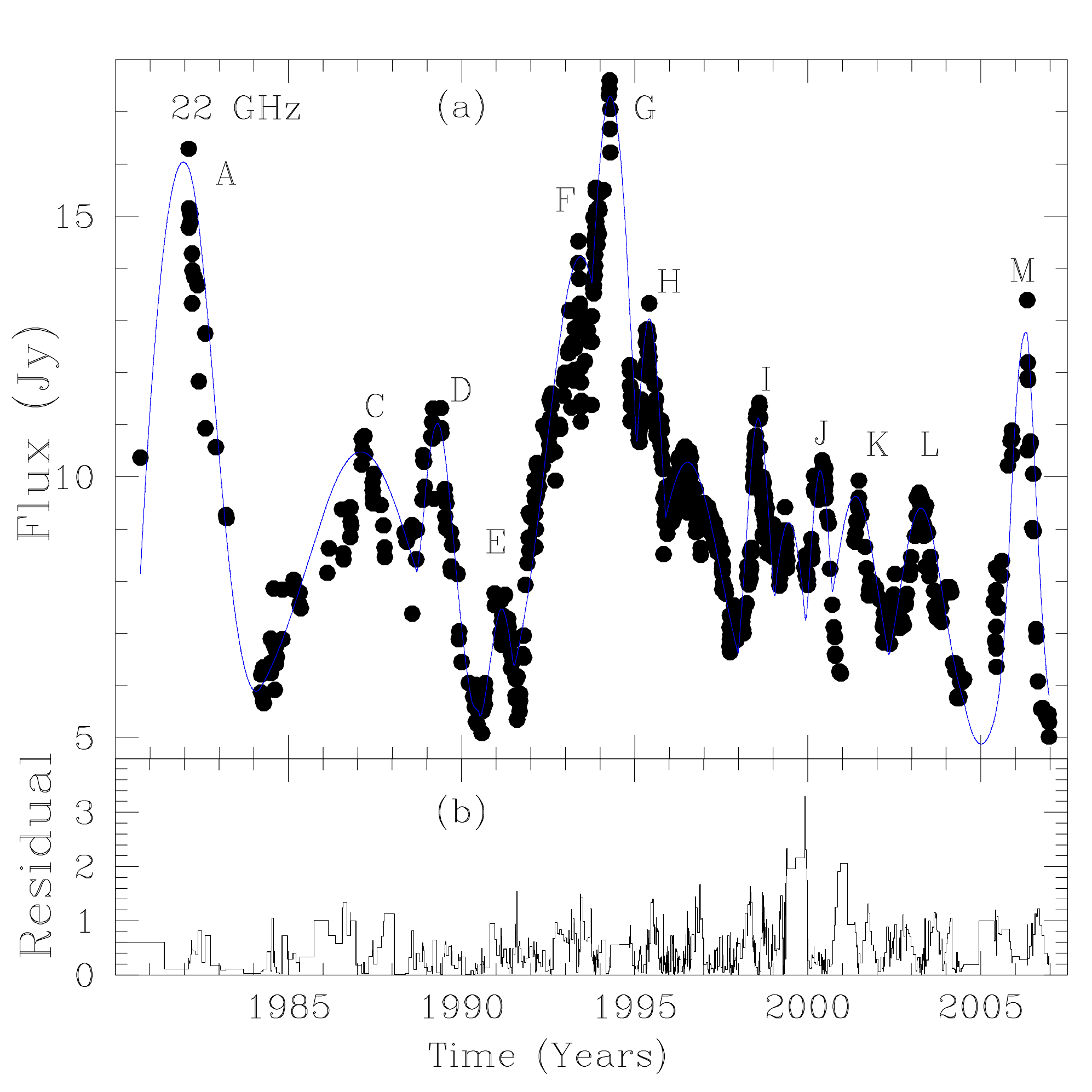}}
\centerline{\includegraphics[scale=0.33]{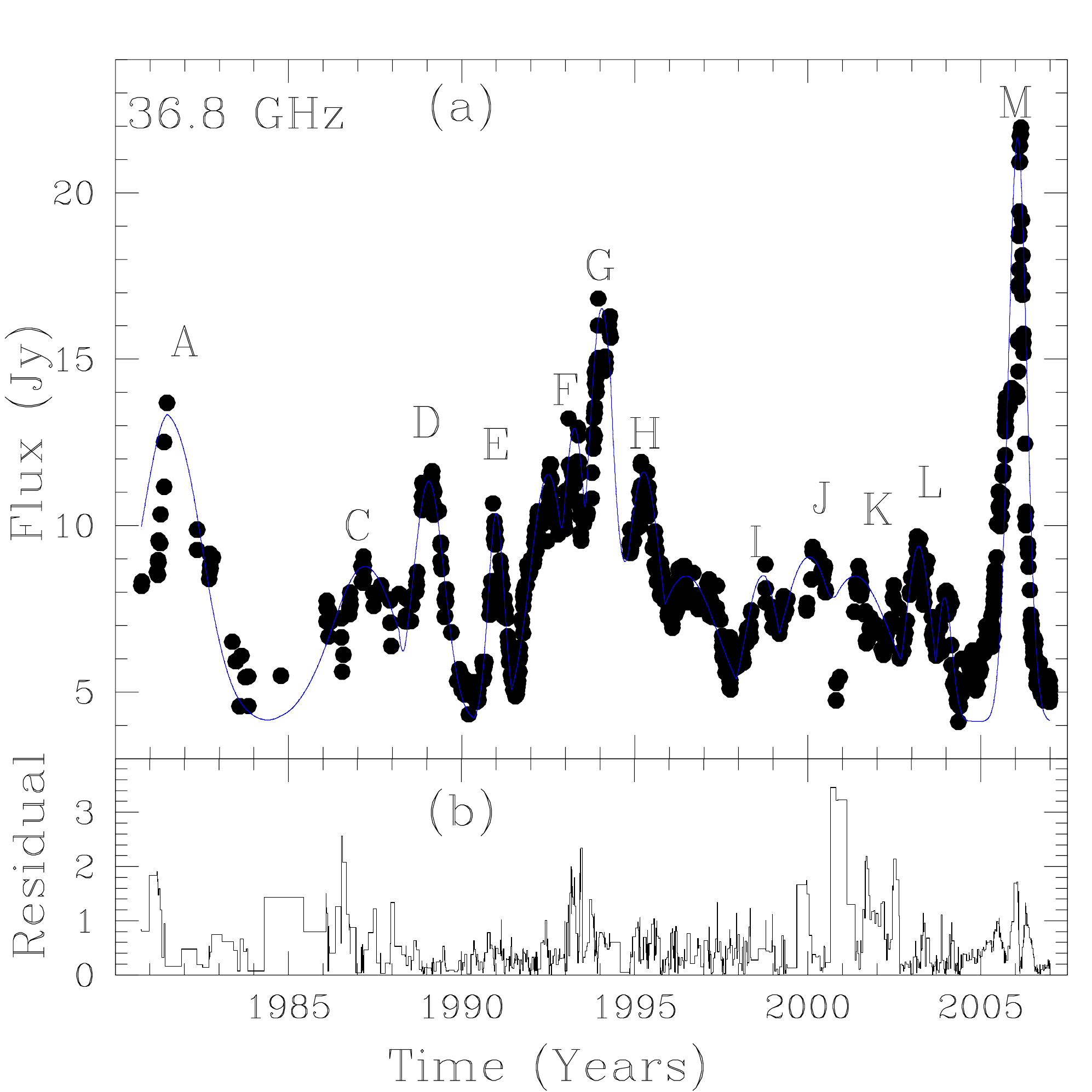}}
\caption{Segment 1 (beginning of observations to $\sim$ 2007.0) of the 22.0 GHz and 36.8 GHz light curve flares fit with a piecewise Gaussian function. The residual in the lower panels are calculated as $[y(A_i,\overline{m}_i,\sigma_i)-x_i(t_i)]/{\rm Standard ~Deviation(x_i (t_i))}$.}
\label{seg1gb}
\end{figure}

\begin{figure}
\centerline{\includegraphics[scale=0.33]{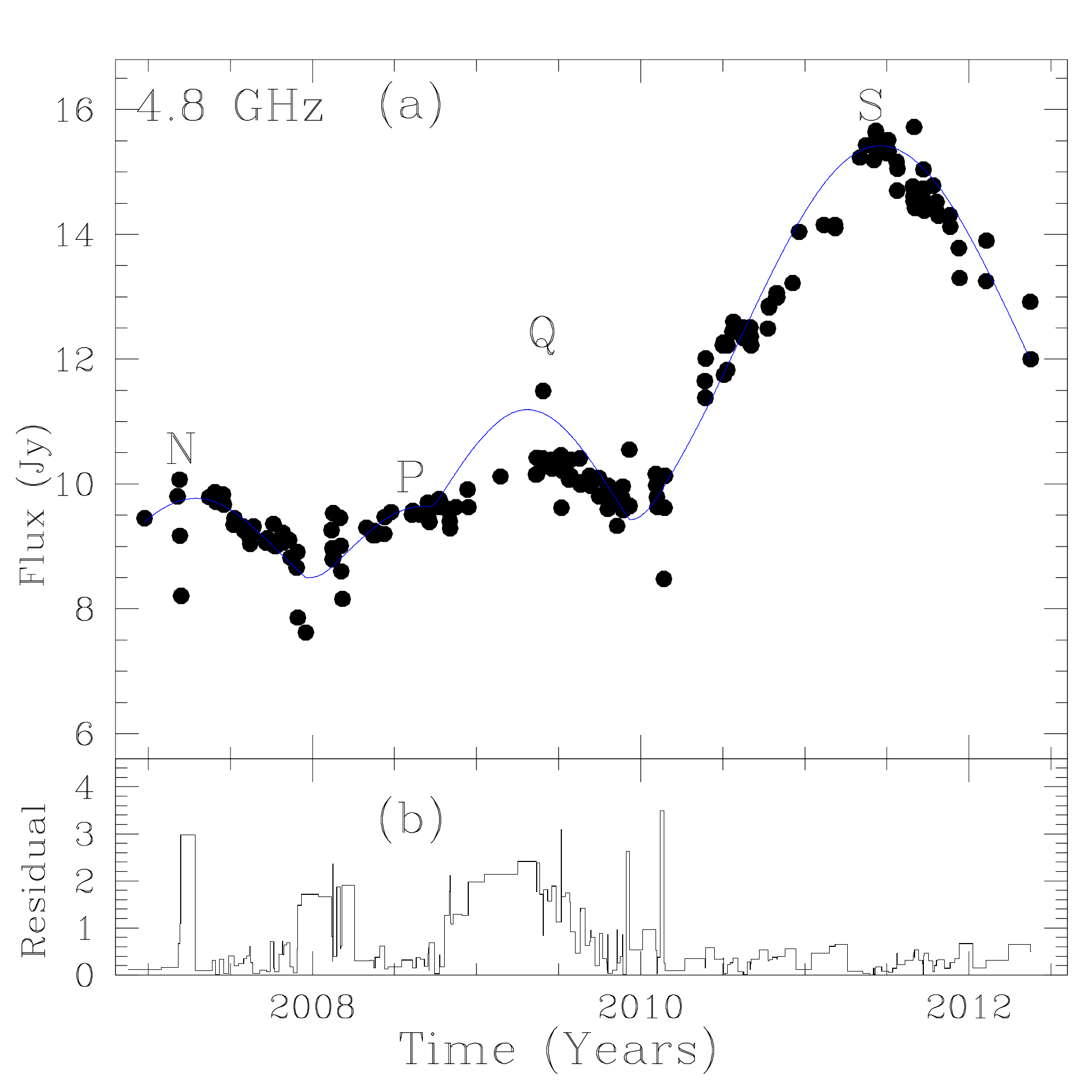}}
\centerline{\includegraphics[scale=0.33]{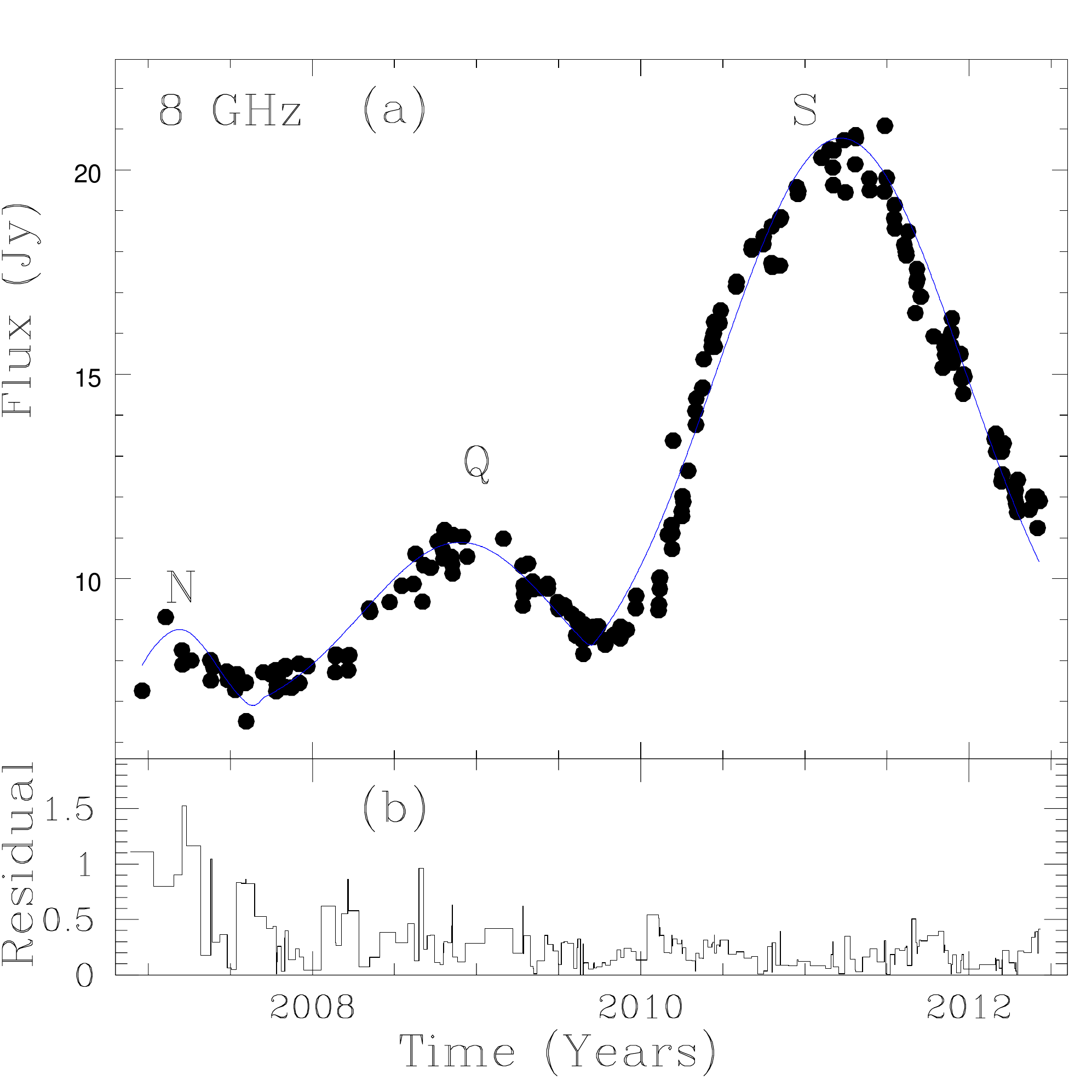}}
\centerline{\includegraphics[scale=0.33]{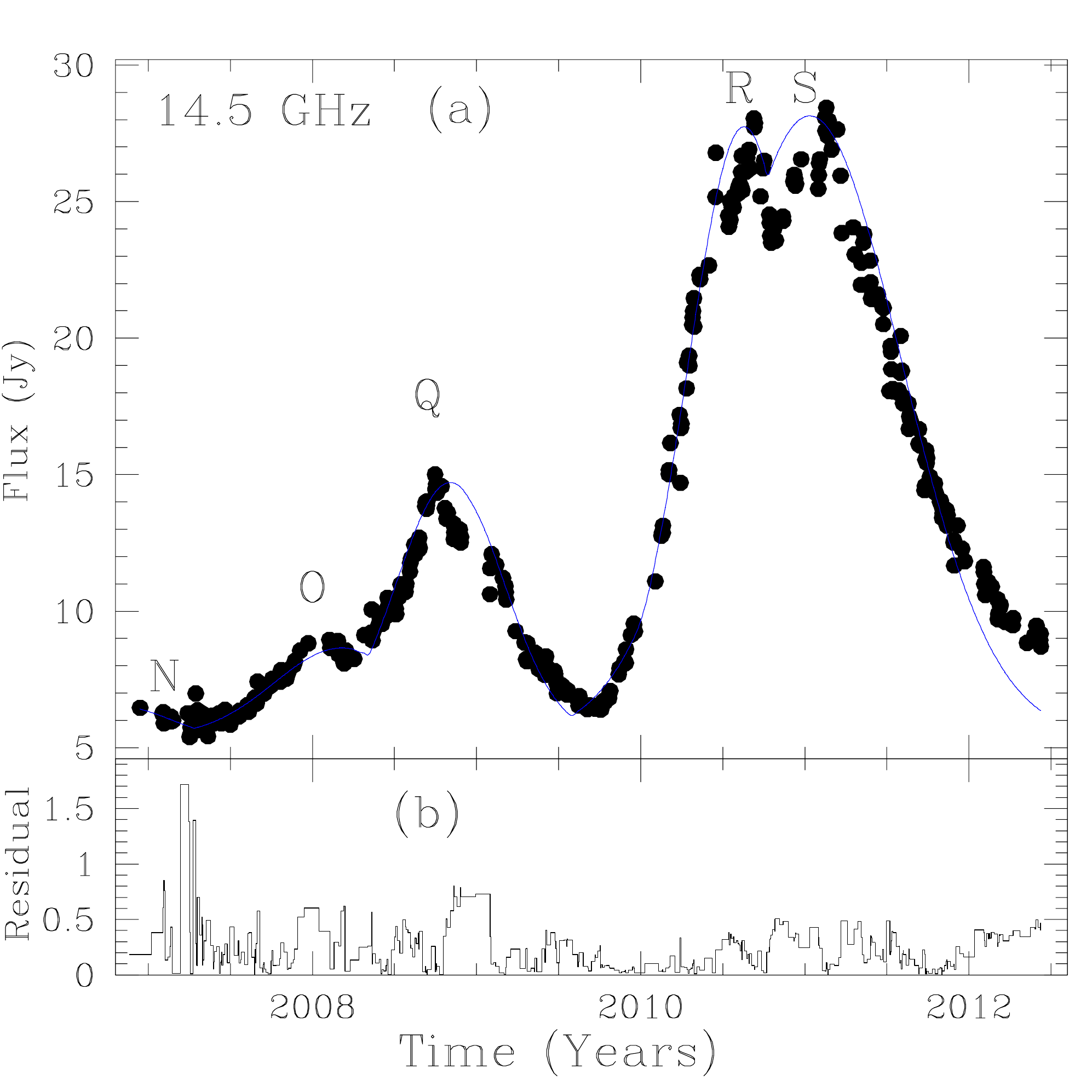}}
\caption{Segment 2 (2007.0 to end of observations) of the 4.8 GHz, 8.0 GHz and 14.5 GHz light curve flares fit with a piecewise Gaussian function. The residual in the lower panels are calculated as $[y(A_i,\overline{m}_i,\sigma_i)-x_i(t_i)]/{\rm Standard ~Deviation(x_i (t_i))}$.}
\label{seg2ga}
\end{figure}

\begin{figure}
\centerline{\includegraphics[scale=0.33]{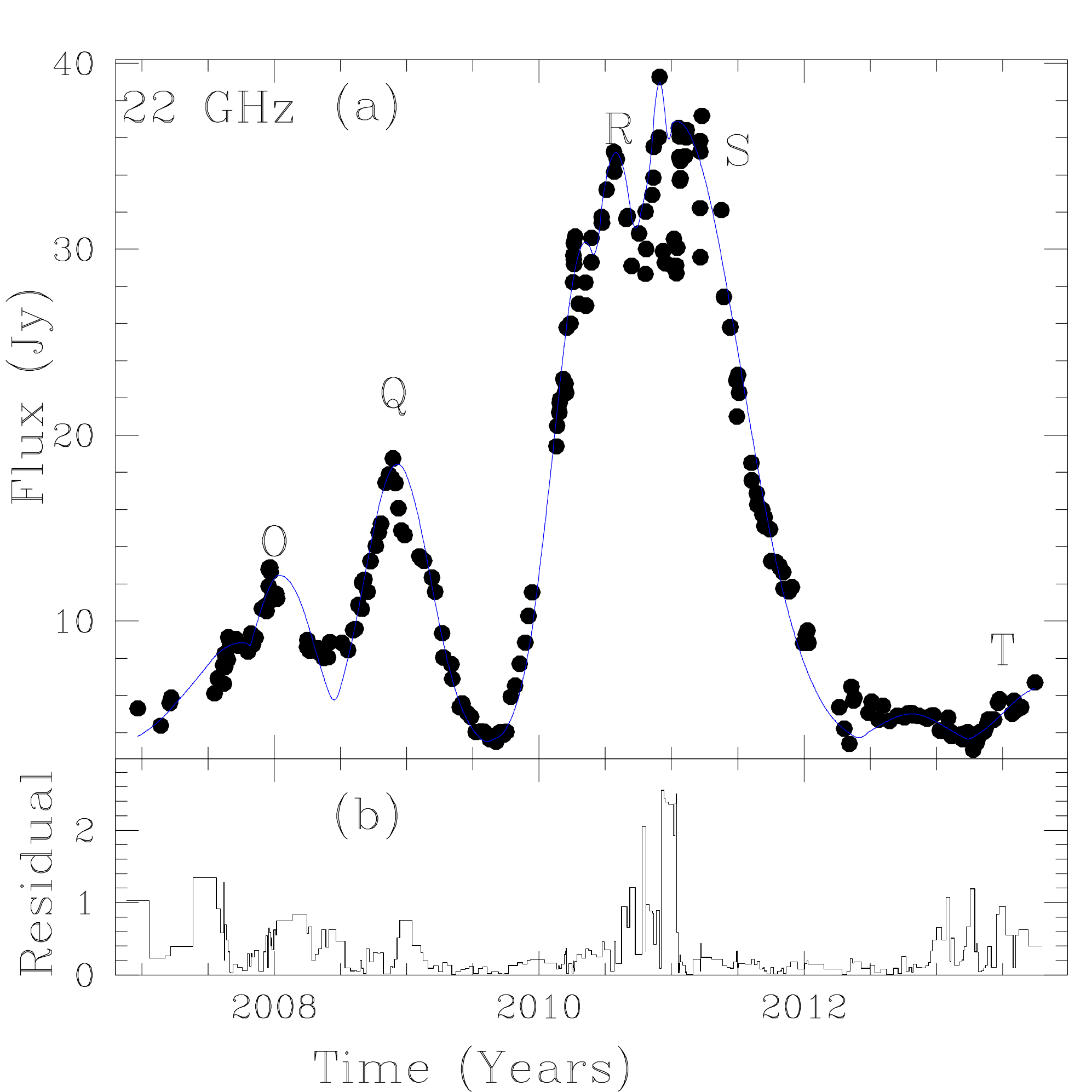}}
\centerline{\includegraphics[scale=0.33]{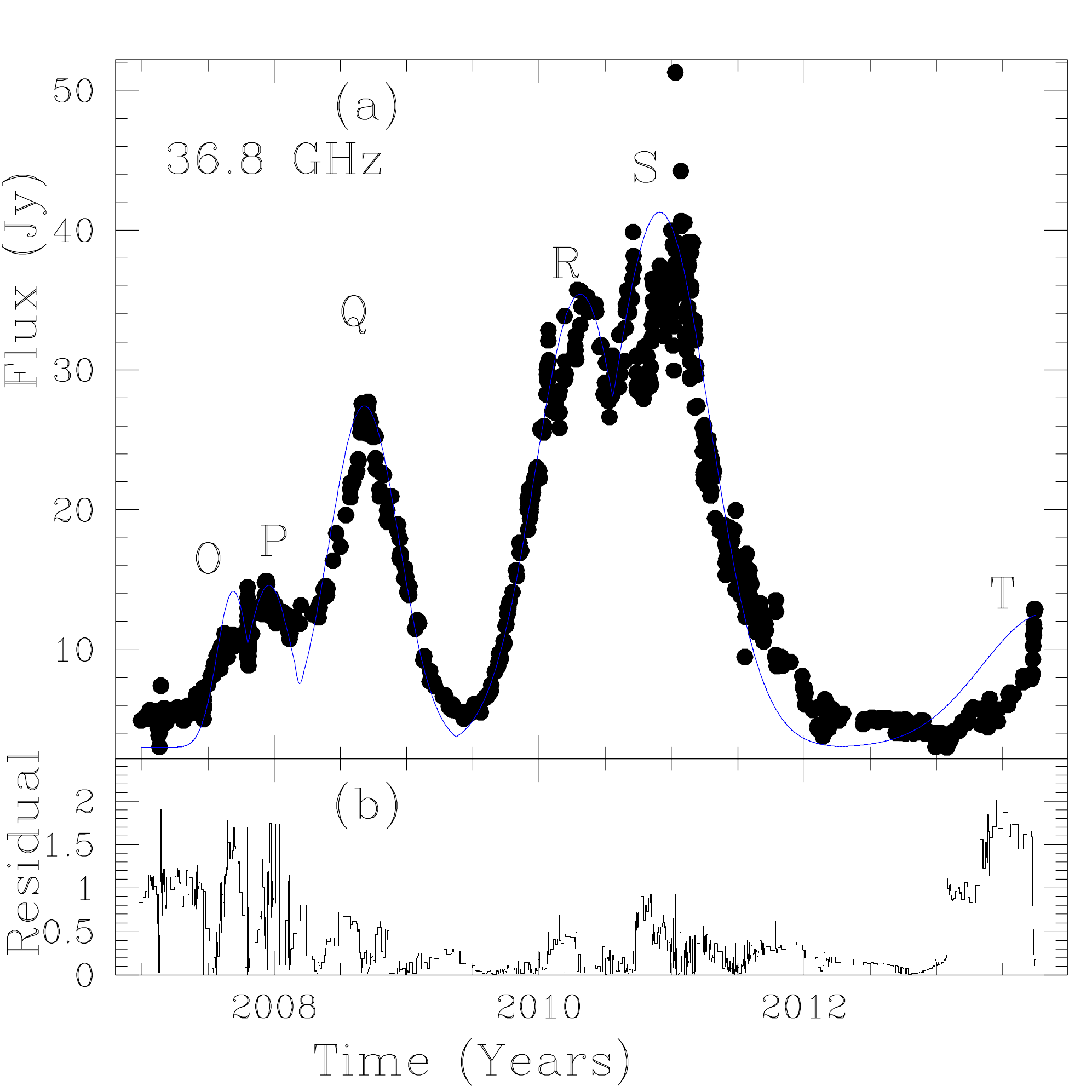}}
\caption{Segment 2 (2007.0 to end of observations) of the 22.0 GHz and 36.8 GHz light curve flares fit with a piecewise Gaussian function. The residual in the lower panels are calculated as $[y(A_i,\overline{m}_i,\sigma_i)-x_i(t_i)]/{\rm Standard~ Deviation(x_i (t_i))}$.}
\label{seg2gb}
\end{figure}

The single Gaussian fit procedure is highly advantageous as it provides a good computational speed (few to tens of seconds depending on parameter grid size); the possibility of using a finely sampled grid of trial values for $A$, $\overline{m}$ and $\sigma$ with a typical grid size of 40000-50000 combinations; better $\chi^2$ fit values as evident from the data fits and the residuals; ease of procedure in determination of best fit parameters and associated errors. The light curves and the best fit Gaussian summed over each segment along with the fit residuals are presented in Figs. \ref{seg1ga}, \ref{seg1gb}, \ref{seg2ga} and \ref{seg2gb}. The residual given by $[y(A_i,\overline{m}_i,\sigma_i)-x_i(t_i)]/{\rm Standard~ Deviation(x_i (t_i))}$ is also calculated with the best fit $(A_i,\overline{m}_i,\sigma_i)$ for each Gaussian segment and plotted in the lower panels in the above figures.

The light curves were split into two segments, one from the start of the observations to 2007.0 and the other from 2007.0 to the end of the observations. This was done to check for consistency of results between the first segment and the second segment, to determine any evolution in the fit parameters and due to the nature of the flaring in the light curve i.e. the flares before 2007.0 are of lower amplitude compared to the post 2007.0 flares. If the entire light curve is run through the Gaussian fit procedure, it is likely to miss out prominent features in the first segment leading to incorrect results. We then conduct a time series analysis of each light curve segment using the Fourier periodogram analysis. As the periodogram is only applicable to evenly sampled light curves, we performed a linear interpolation and sampled the light curve at regular intervals of 0.1 days, similar to the procedure followed in, e.g.~ \cite{2012A&A...544A..80G,2014ApJ...791...74M}. This evenly sampled light curve is used to determine the power spectral density (PSD) shape and any statistically significant quasi-periodic oscillations (QPOs). The normalized periodogram is given by (e.g. \citealt{2003MNRAS.345.1271V}),
\begin{equation}
P(f_j)=\frac{2 \Delta t}{\overline{x}^2 N} |F(f_j)|^2
\end{equation}
where $\Delta t = 0.1 ~\rm{days}$ is the sampling time step, $\overline{x}$ is the mean of the light curve $x(t_k)$ of length $N$ points, and $|F(f_j)|$ is its discrete Fourier transform evaluated at frequencies $f_j = j/(N \Delta t)$ with  $j = 1, 2,..,(N/2-1)$. We fit the periodogram with two competing models, a power law and a bending power law, in order to describe the PSD shape. The power law model is given by
\begin{equation}
I(f_j) = A f^{s}_j+C,
\end{equation}
with amplitude $A$, slope $s$, and a constant Poisson noise $C$. This model fits optical/ultra-violet and X-ray data reasonably well as inferred from previous studies (e.g. \citealt{2012A&A...544A..80G,2014ApJ...791...74M}) and could thus represent broadband variability across a wide range of Fourier frequencies in multiple wavelengths. The bending power law model is given by
\begin{eqnarray}
I(f_j) = A f^{-1}_j \left(1+(f_j/f_b)^{-s-1}\right)^{-1}+C.
\end{eqnarray}
with amplitude $A$, slope $s$, bend frequency $f_{b}$, and a constant Poisson noise $C$. The model is a special case of the generalized bending knee model \citep{2004MNRAS.348..783M}. The maximum likelihood estimator (MLE) method is used to determine model parameters in which the log-likelihood function (e.g. \citealt{2013MNRAS.433..907E,2014JApA..tmp...75M,2014ApJ...791...74M}) defined by
\begin{equation}
 S(\theta_k) = - 2 \sum^{n-1}_{j=1}(\ln(I(f_j,\theta_k))+P(f_j)/I(f_j,\theta_k)),
\end{equation}
is first determined. Here, $I(f_j,\theta_k)$ are the power law or bending power model with parameters $\theta_k$. Determining $\theta_k$ which minimize $S$ yields the maximum likelihood values. The log-likelihood $S$ is determined for a large number of combinations of the parameters $\theta_k$ for a given model. The parameter combination which gives a unique global minimum $S_{\mathrm{min}}$ yields the best fit. For $S_i$ corresponding to each unique combination of the parameters $\theta_k$, as $\Delta S= S_i - S_{\mathrm{min}}$ are approximately $\chi^2_k$ distributed, the cumulative distribution function of the $\chi^2_k$ distribution is used to estimate parameter confidence intervals. The $\Delta S$ determined from the parameter combinations within a desired confidence interval are grouped. Parameter ranges that they correspond to are used to determine confidence intervals of the $\theta_k$ used in a given model.

For model selection, we use the Akaike information criteria (AIC) (e.g. \citealt{2004..Likelihood}) which measures the information lost when a model is fit to the data. The model with least information loss (least AIC) is the best fit. A likelihood function $L$ which is proportional to the probability of the success of the model in describing the PSD shape is evaluated for each model. The AIC and likelihood are defined by
\begin{align}
AIC&=S(\theta_k)+2 p_k, \\ \nonumber
\Delta_i&=AIC_{\mathrm{min (model \ i)}}-AIC_{\mathrm{min(null)}}, \\ \nonumber
L(\mathrm{model \ i|data})&=e^{-\Delta_i/2},
\end{align}
where $p_k$, the penalty term,  is the number of parameters $\theta_k$ used in the model, and $L(\mathrm{model \ i|data})$ is the likelihood of model $i$ given the data. Models with $\Delta_i\leq 2$ can be considered close to the best fit, those with $4 \leq \Delta_i \leq 7$ are considerably less supported, and those with $\Delta_i > 10$ cannot be supported \citep{2004..Likelihood}. As the residual $\gamma(f_j) = P(f_j)/I(f_j)$ follows the $\chi^2_2$ distribution, the area under the tail of the probability density function of the $\chi^2_2$ distribution (gamma density $\Gamma(1,1/2)= \exp{(-x/2)}/2$) gives the probability $\epsilon$ that the power deviates from the mean at a given frequency and is measured in units of standard deviation given by $\gamma_{\epsilon}$.  We must correct $\gamma_{\epsilon}$  to account for the $K$ number of trial frequencies at which the periodogram is evaluated and so is given by (\citealt{2005A&A...431..391V}),
\begin{equation}
\gamma_\epsilon=-2 \ln[1-(1-\epsilon)^{\frac{1}{K}}].
\end{equation}
Once $\epsilon$ (e.g. 0.95, 0.99) is specified, $\gamma_\epsilon$ is calculated and multiplied with $I(f_j)$ to give the significance level used to identify outliers in the periodogram that could indicate the presence of  a QPO.

\section{Results and discussion}
\label{results}

For a flare $X$ observed in multiple frequencies $\nu$, $\overline{m}_{X,\nu_1}-\overline{m}_{X,\nu_2} = \Delta t_X$ where $\Delta t_X$ is the time lag between the flare in $\nu_1$ and $\nu_2$ with $\nu_1>\nu_2$. The parameters derived from the Gaussian fit procedure discussed in section \ref{observations} are presented in Table \ref{tab1} for the first segment and in Table \ref{tab2} for the second segment. In both tables, column 1 presents the flare nomenclature; column 2 gives the observed frequency, column 3 gives the maximum amplitude of the flare at a particular frequency; column 4 gives the epoch of the maximum flare amplitude; column 5 gives the full width at half maxima obtained from Gaussian fit for the particular flare; column 6 gives the calculated time lags and column 7 gives the spectral index for each flare.

\begin{figure}
\centerline{\includegraphics[scale=0.43]{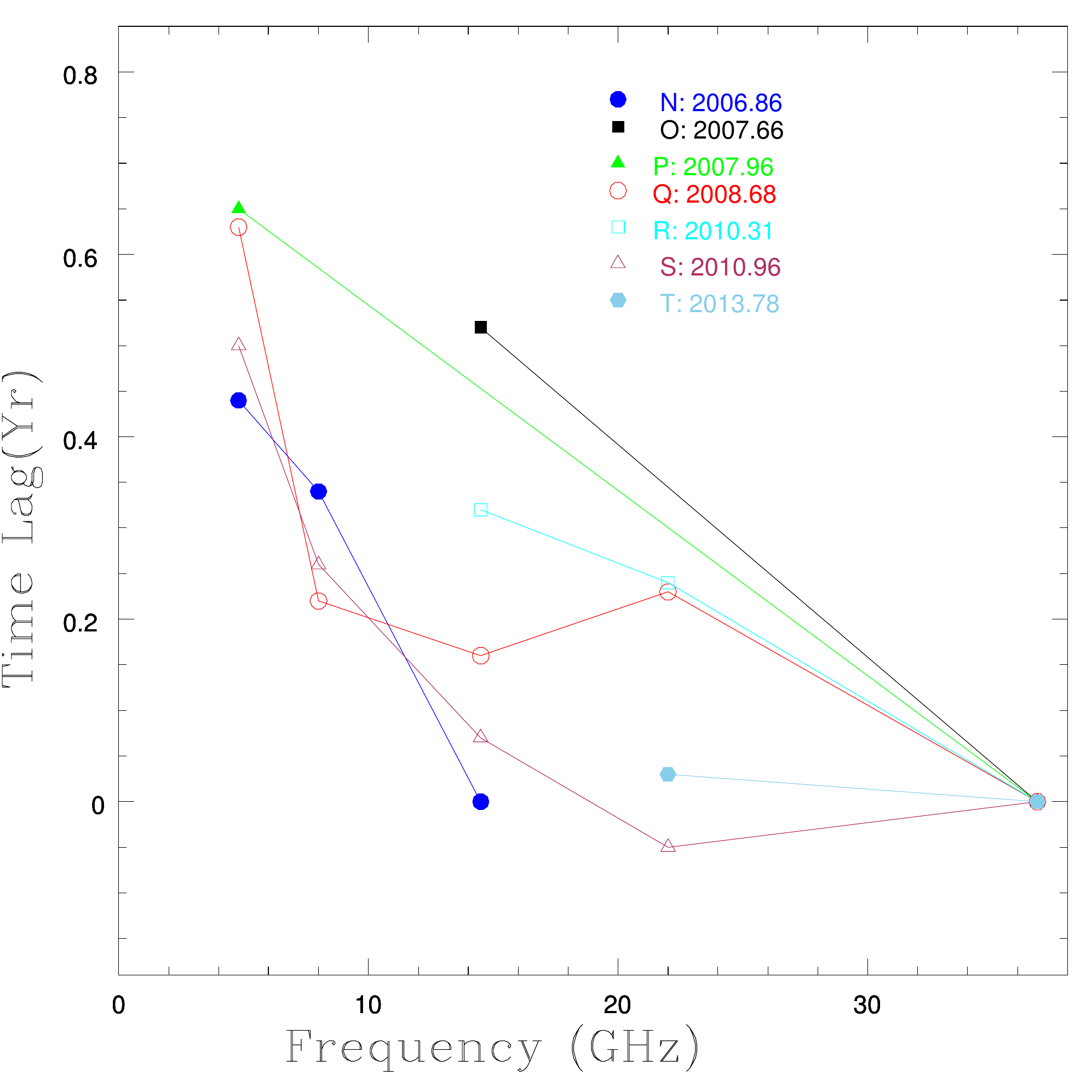}}
\caption{Time lags $\Delta t$ as a function of $\nu$ for segment 2 light curves. A fit to $\Delta t = a \nu^{-1/k_r}+b$ yields smaller magnitudes for $b = -0.06 \pm 0.02$ yr and $a = 3.03 \pm 0.89$ yr (GHz)$^{-1/k_r}$ and a consistent $k_r = 1.22 \pm 0.33 $ when compared with the fit to segment 1 light curves. 
}
\label{tnu1plot}
\end{figure}

We fit time lags ($\Delta t_X$) derived from the Gaussian fits described in section \ref{observations} with a function $\Delta t = a \nu^{-1/k_r}+b$ using a weighted non-linear least squares method. The median of each parameter and the associated median deviation are reported below. Those portions that were not fit well due to either too small a number of points or unstable numerical results are excluded in the calculation of the parameters. The results of $\Delta t$ versus $\nu$ are plotted in Fig. \ref{tnuplot} for segment 1 and Fig. \ref{tnu1plot} for segment 2 and the values of the fit parameters are: 
\begin{figure}
\centerline{\includegraphics[scale=0.43]{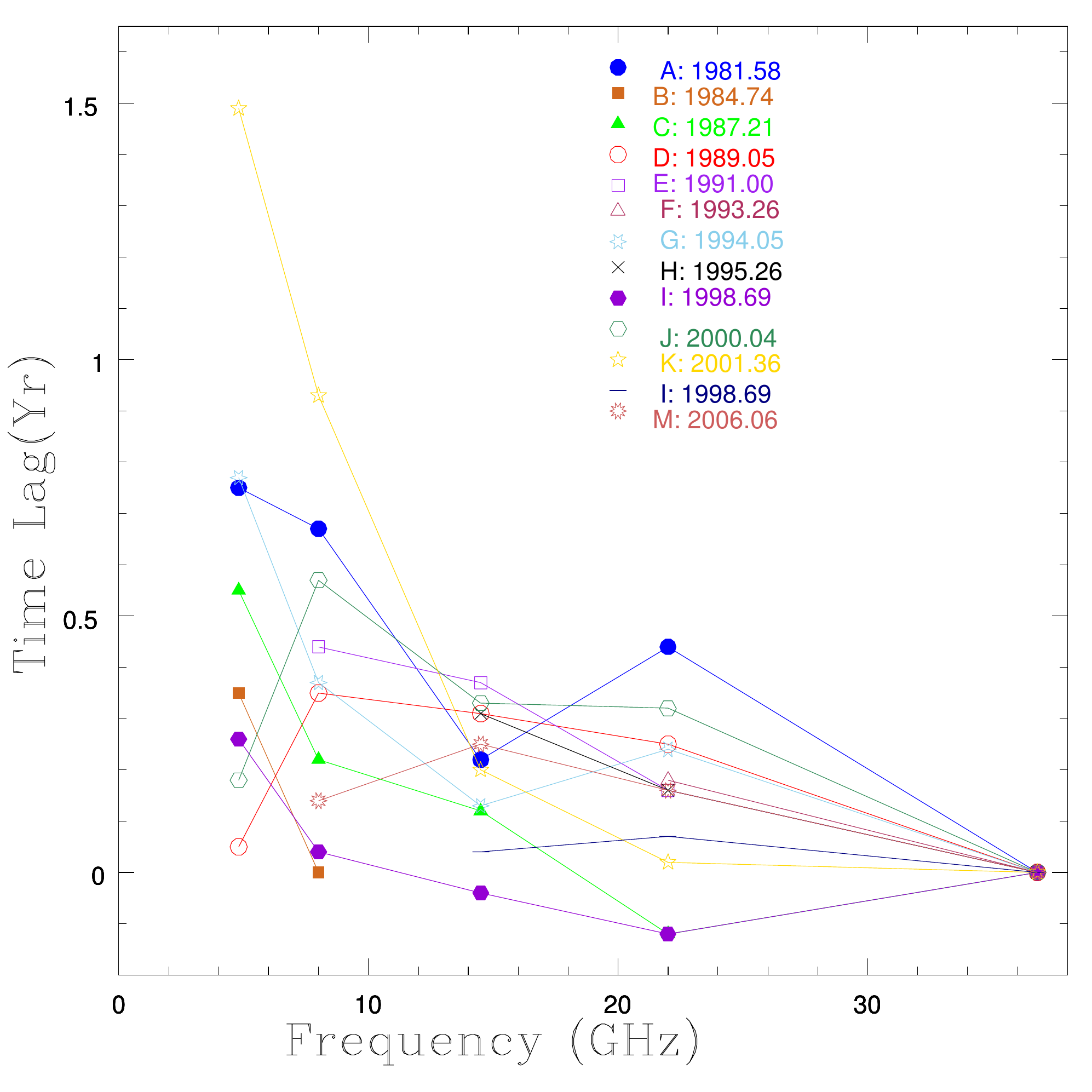}}
\caption{Time lags $\Delta t$ as a function of $\nu$ for segment 1 light curves. A fit to $\Delta t = a \nu^{-1/k_r}+b$ yields $b = -0.28 \pm 0.19$ yr, $a = 5.51 \pm 1.03$ yr (GHz)$^{-1/k_r}$ and $k_r = 0.97 \pm 0.24$. The $k_r$ value indicates consistency of the equipartition between magnetic field energy density and the particle kinetic energy.}
\label{tnuplot}
\end{figure}

\begin{enumerate}
\item First segment: beginning of observation -- 2007.0
\begin{align*}
b &= -0.28 \pm 0.19 ~ yr\\
a &= 5.51 \pm 1.03 ~ yr~(GHz)^{-1/k_r}\\
k_r &= 0.97 \pm 0.24 ;
\end{align*} 
\item Second segment: 2007.0 -- end of observation
\begin{align*}
b &= -0.06 \pm 0.02 ~ yr\\
a &= 3.03 \pm 0.87 ~ yr~(GHz)^{-1/k_r}\\
k_r &= 1.22 \pm 0.33 .
\end{align*}
\end{enumerate}

In both light curve segments, the inferred $k_r$ are similar and consistent with equipartition between particle energy density and magnetic field energy density. From the above fits, we have information on the amplitude and hence the peak flux for each flare at the observed frequencies. As $S \propto \nu^{\alpha}$, the amplitude versus frequency data is fit with a function $A = N~ \nu^{\alpha}$ to determine the spectral index $\alpha$ for each flare. The fit is carried out only for flares with four data points or more. We find $\alpha$ in the range $-0.24$ and $1.52$ for segment 1 and $1.16$ and $1.31$ for flares Q and S of segment 2. The calculated $\alpha$ are presented in Table \ref{tab1} for segment 1 and in Table \ref{tab2} for segment 2. In a majority of cases (7/11), we infer a positive $\alpha$ indicating that there is in general an increase in amplitude with frequency for that flare. For two flares (C and K), we obtain the opposite trend, i.e. decreasing amplitude with increasing frequency, while for the remaining two (A and I) there is no significant dependence.

As the equipartition condition is satisfied in both segments, we use eqns. (\ref{B1}) and (\ref{Bcore}) to determine the magnetic field strength at 1 pc from the jet base and at the core respectively; $B_1$, $B_{\rm core}$, the core position offset from eqn. (\ref{omega}) and the distance of the VLBI core from the base of the jet from eqn. (\ref{rcore}) are calculated and presented in Table \ref{tab3} for the first segment and Table \ref{tab4} for the second segment. Using $\theta = 1.3^{\circ}$, $\phi = 0.8^{\circ}$, $\delta = 24.6$, average proper motion $\mu = 0.12$ mas/yr (\citealt{2010ApJ...715..362J}), $z = 0.859~$ and $D = 5.489$ Gpc ($\Omega_\Lambda = 0.73$ for the vacuum energy density, $\Omega_{m} = 0.27$ for the matter energy density and Hubble constant $H_0 = 71~$km/s/Mpc), we obtain weighted mean values of $B_1 = 0.49 \pm 0.18$ G for segment 1, $B_1 = 0.36 \pm 0.05$ G for segment 2, $B_{\rm core} = 47 \pm 16$ mG for segment 1 and $B_{\rm core} = 23 \pm 7$ mG for segment 2. We obtain a weighted mean $\Omega_{r\nu} = 6.59 \pm 3.09$ pc GHz$^{1/k_r}$ when averaged over all frequency pairs in the first segment and $\Omega_{r\nu} = 4.25 \pm 0.78$ pc GHz$^{1/k_r}$ in the second segment. The core radius $r_{\rm core}$ is expected to decrease with increasing $\nu$. For 9 out of 20 flares, namely, B, F, H, L, N, O, P, R, and T, no conclusion can be drawn as the number of data points is two or one. The expected trend is inferred in 7 of the remaining 11 flares with the removal of one of the data points, the 4.8 GHz (flare K), 14.5 GHz (flare E) and 22 GHz (flares A, C, G, I and S). The remaining 4 flares (D, J, M and Q) show the reverse trend. This indicates that the synchrotron opacity model may not be valid in all cases. In both tables, column 1 presents the flare nomenclature; column 2 gives the observed frequency, column 3 gives the core position offset in mas; column 4 gives the core position offset in pcGHz; column 5 gives the distance of the emitting core from the base of the jet; column 6 gives the magnetic field strength in Gauss at 1 pc and column 7 gives the magnetic field strength in Gauss at the emitting core.

\begin{table}
\centering
\scalebox{.73}{
\begin{tabular}{|l|l|l|l|l|l|l|}
\hline
Flare & Frequency & Amplitude & Position & Width & Time lag & Spectral index \\
      & (GHz)     & $A$ (Jy)  & $\overline{m}$ (year) & $\sigma$ (years) & $\Delta t$ (years) & $\alpha$ \\
\hline
A     & 36.8      & 9.28$\pm$ 0.40 & 1981.58$\pm$0.08 & 0.86$\pm$0.19 & 0.00 & $-$0.09$\pm$0.13\\
      & 22.0      & 10.97$\pm$0.18 & 1982.02$\pm$0.14 & 0.82$\pm$0.39 & 0.44$\pm$0.16 & \\
      & 14.5      & 12.63$\pm$0.06 & 1981.80$\pm$0.10 & 0.96$\pm$0.51 & 0.22$\pm$0.17 & \\
      & 8.0       & 12.38$\pm$0.30 & 1982.25$\pm$0.09 & 0.96$\pm$0.36 & 0.67$\pm$0.13 & \\
      & 4.8       & 9.01$\pm$0.54  & 1982.33$\pm$0.17 & 1.02$\pm$0.15 & 0.75$\pm$0.19 & \\ \hline
B     & 8.0       & 2.90$\pm$0.46 & 1984.74$\pm$0.14 & 0.39$\pm$0.06 & 0.00 & \\
      & 4.8       & 3.39$\pm$0.10 & 1985.09$\pm$0.16 & 0.55$\pm$0.11 & 0.35$\pm$0.21 & \\ \hline
C     & 36.8      & 4.66$\pm$0.36 & 1987.21$\pm$0.02 & 0.95$\pm$0.33 & 0.00 & -0.19 $\pm$ 0.13\\ 
      & 22.0      & 5.46$\pm$0.28 & 1987.09$\pm$0.13 & 1.54$\pm$0.43 & -0.12$\pm$0.13 & \\
      & 14.5      & 7.16$\pm$0.02 & 1987.33$\pm$0.06 & 1.03$\pm$0.50 & 0.12$\pm$0.14 & \\
      & 8.0       & 7.62$\pm$0.42 & 1987.43$\pm$0.03 & 1.17$\pm$0.25 & 0.22$\pm$0.07 & \\
      & 4.8       & 6.98$\pm$0.46 & 1987.76$\pm$0.06 & 1.33$\pm$0.27 & 0.55$\pm$0.07 & \\ \hline
D     & 36.8      & 7.22$\pm$0.28 & 1989.05$\pm$0.02 & 0.45$\pm$0.27 & 0.00 & 0.12 $\pm$ 0.02 \\
      & 22.0      & 6.00$\pm$0.38 & 1989.30$\pm$0.18 & 0.48$\pm$0.04 & 0.25$\pm$0.18 & \\
      & 14.5      & 6.37$\pm$0.42 & 1989.36$\pm$0.13 & 0.55$\pm$0.16 & 0.31$\pm$0.22 & \\
      & 8.0       & 6.21$\pm$0.52 & 1989.40$\pm$0.03 & 0.59$\pm$0.18 & 0.35$\pm$0.13 & \\
      & 4.8       & 5.47$\pm$0.14 & 1989.10$\pm$0.01 & 1.09$\pm$0.30 & 0.05$\pm$0.03 & \\ \hline
E     & 36.8      & 6.26$\pm$0.56 & 1991.01$\pm$0.09 & 0.23$\pm$0.12 & 0.00 & 1.52 $\pm$ 0.23 \\
      & 22.0      & 2.45$\pm$0.24 & 1991.16$\pm$0.14 & 0.31$\pm$0.08 & 0.16$\pm$0.17 & \\
      & 14.5      & 1.67$\pm$0.32 & 1991.37$\pm$0.15 & 0.48$\pm$0.10 & 0.37$\pm$0.21 & \\
      & 8.0       & 0.86$\pm$0.28 & 1991.44$\pm$0.03 & 0.13$\pm$0.03 & 0.44$\pm$0.15 & \\ \hline
F     & 36.0      & 8.81$\pm$0.08 & 1993.26$\pm$0.05 & 0.36$\pm$0.15 & 0.00 & \\
      & 22.0      & 9.2$\pm$0.02  & 1993.44$\pm$0.13 & 0.98$\pm$0.31 & 0.18$\pm$0.14 & \\ \hline
G     & 36.8      & 12.41$\pm$0.01 & 1994.05$\pm$0.15 & 0.41$\pm$0.14 & 0.00 & 0.16 $\pm$ 0.01\\
      & 22.0      & 12.28$\pm$0.30 & 1994.29$\pm$0.08 & 0.63$\pm$0.17 & 0.24$\pm$0.17 & \\
      & 14.5      & 15.61$\pm$0.24 & 1994.18$\pm$0.02 & 0.92$\pm$0.25 & 0.13$\pm$0.08 & \\
      & 8.0       & 13.37$\pm$0.30 & 1994.42$\pm$0.16 & 0.77$\pm$0.27 & 0.37$\pm$0.16 & \\
      & 4.8       & 8.94$\pm$0.01 & 1994.82$\pm$0.11 & 1.06$\pm$0.37 & 0.77$\pm$0.19 & \\ \hline
H     & 36.8      & 7.50$\pm$0.10 & 1995.26$\pm$0.18 & 0.50$\pm$0.01 & 0.00 & \\ 
      & 22.0      & 8.01$\pm$0.42 & 1995.42$\pm$0.07 & 0.42$\pm$0.07 & 0.16$\pm$0.19 & \\
      & 14.5      & 7.87$\pm$0.20 & 1995.57$\pm$0.19 & 0.37$\pm$0.05 & 0.31$\pm$0.20 & \\ \hline
I     & 36.8      & 4.43$\pm$0.34 & 1998.69$\pm$0.10 & 0.48$\pm$0.17 & 0.00 & -0.04 $\pm$ 0.09\\
      & 22.0      & 6.10$\pm$0.46 & 1998.57$\pm$0.09 & 0.36$\pm$0.09 & -0.12$\pm$0.13 & \\
      & 14.5      & 6.14$\pm$0.26 & 1998.65$\pm$0.16 & 0.41$\pm$0.06 & -0.04$\pm$0.18 & \\
      & 8.0       & 5.60$\pm$0.14 & 1998.73$\pm$0.17 & 1.29$\pm$0.34 & 0.04$\pm$0.23 & \\
      & 4.8       & 4.79$\pm$0.52 & 1998.95$\pm$0.06 & 1.22$\pm$0.38 & 0.26$\pm$0.18 & \\ \hline
J     & 36.8      & 4.95$\pm$0.12 & 2000.04$\pm$0.14 & 0.77$\pm$0.15 & 0.00 & 0.08 $\pm$ 0.06\\
      & 22.0      & 5.10$\pm$0.08 & 2000.36$\pm$0.05 & 0.31$\pm$0.13 & 0.32$\pm$0.15 & \\
      & 14.5      & 4.60$\pm$0.12 & 2000.37$\pm$0.13 & 0.91$\pm$0.27 & 0.33$\pm$0.14 & \\
      & 8.0       & 3.61$\pm$0.58 & 2000.61$\pm$0.06 & 0.28$\pm$0.10 & 0.57$\pm$0.14 & \\
      & 4.8       & 4.41$\pm$0.38 & 2001.22$\pm$0.09 & 0.78$\pm$0.26 & 0.18$\pm$0.11 & \\ \hline
K     & 36.8      & 4.37$\pm$0.50 & 2001.36$\pm$0.12 & 1.01$\pm$0.18 & 0.00 & -0.24 $\pm$ 0.17\\
      & 22.0      & 4.61$\pm$0.04 & 2001.38$\pm$0.14 & 0.65$\pm$0.11 & 0.02$\pm$0.18 & \\
      & 14.5      & 5.52$\pm$0.10 & 2001.56$\pm$0.03 & 0.49$\pm$0.16 & 0.20$\pm$0.14 & \\
      & 8.0       & 4.20$\pm$0.36 & 2002.29$\pm$0.20 & 1.46$\pm$0.37 & 0.93$\pm$0.20 & \\ 
      & 4.8       & 4.21$\pm$0.58 & 2002.85$\pm$0.01 & 1.23$\pm$0.31 & 1.49$\pm$0.20 & \\ \hline
L     & 36.8      & 5.27$\pm$0.10 & 2003.20$\pm$0.09 & 0.35$\pm$0.08 & 0.00 & \\
      & 22.0      & 4.38$\pm$0.20 & 2003.27$\pm$0.12 & 0.64$\pm$0.16 & 0.07$\pm$0.15 & \\
      & 14.5      & 5.05$\pm$0.36 & 2003.24$\pm$0.01 & 0.77$\pm$0.17 & 0.04$\pm$0.12 & \\ \hline
M     & 36.8      & 17.55$\pm$0.02 & 2006.06$\pm$0.03 & 0.26$\pm$0.37 & 0.00 & 1.05 $\pm $0.06\\
      & 22.0      & 8.07$\pm$0.44 & 2006.24$\pm$0.18 & 0.34$\pm$0.03 & 0.16$\pm$0.18 & \\
      & 14.5      & 5.08$\pm$0.50 & 2006.35$\pm$0.05 & 1.09$\pm$0.15 & 0.25$\pm$0.19 & \\
      & 8.0       & 4.38$\pm$0.22 & 2006.20$\pm$0.01 & 0.58$\pm$0.45 & 0.14$\pm$0.05 & \\
\hline
\end{tabular}}
\caption{Gaussian fit based parameters, time lags and spectral index for segment 1 light curves.}
\label{tab1}
\end{table}

\begin{table}
\centering
\scalebox{.73}{
\begin{tabular}{|l|l|l|l|l|l|l|}
\hline
Flare & Frequency & Amplitude & Position & Width & Time lag & Spectral index \\
      & (GHz)     & $A$ (Jy)  & $\overline{m}$ (year) & $\sigma$ (years) & $\Delta t$ (years) & $\alpha$ \\
\hline
N     & 14.5      & 1.08$\pm$0.12 & 2006.85$\pm$0.08 & 0.28$\pm$0.12 & 0.00 &\\
      & 8.0       & 2.25$\pm$0.54 & 2007.19$\pm$0.01 & 0.23$\pm$0.06 & 0.34$\pm$0.08 & \\
      & 4.8       & 2.15$\pm$0.56 & 2007.29$\pm$0.13 & 0.50$\pm$0.15 & 0.44$\pm$0.13 & \\ \hline
O     & 36.0      & 11.19$\pm$ 0.22 & 2007.66$\pm$0.01 & 0.15$\pm$0.05 & 0.00 &  \\
      & 22.0      & 9.50$\pm$0.28 & 2008.06$\pm$0.12 & 0.23$\pm$0.02 & 0.40$\pm$0.12 & \\
      & 14.5      & 3.25$\pm$0.36 & 2008.18$\pm$0.14 & 0.41$\pm$0.20 & 0.52$\pm$0.14 & \\ \hline
P     & 36.0      & 11.61$\pm$ 0.46 & 2007.96$\pm$0.06 & 0.17$\pm$0.03 & 0.00 & \\
      & 4.8       & 1.98$\pm$0.24 & 2008.61$\pm$0.23 & 0.48$\pm$0.04 & 0.65$\pm$0.24 & \\ \hline
Q     & 36.0      & 24.44$\pm$ 0.60 & 2008.68$\pm$0.07 & 0.26$\pm$0.09 & 0.00 & 1.16 $\pm$ 0.06 \\
      & 22.0      & 15.36$\pm$0.06 & 2008.92$\pm$0.04 & 0.27$\pm$0.26 & 0.24$\pm$0.08 & \\
      & 14.5      & 9.31$\pm$0.18 & 2008.84$\pm$0.04 & 0.33$\pm$0.14 & 0.16$\pm$0.04 & \\
      & 8.0       & 4.38$\pm$0.08 & 2008.90$\pm$0.16 & 0.60$\pm$0.21 & 0.22$\pm$0.16 & \\
      & 4.8       & 3.57$\pm$0.54  & 2009.31$\pm$0.09 & 1.02$\pm$0.15 & 0.53$\pm$0.10 & \\ \hline
R     & 36.0      & 32.43$\pm$ 0.24 & 2010.31$\pm$0.02 & 0.34$\pm$0.09 & 0.00 &  \\
      & 22.0      & 31.85$\pm$0.34 & 2010.54$\pm$0.01 & 0.36$\pm$0.12 & 0.23$\pm$0.02 & \\
      & 14.5      & 22.34$\pm$0.06 & 2010.63$\pm$0.07 & 0.34$\pm$0.15 & 0.32$\pm$0.07 & \\ \hline
S     & 36.0      & 47.99$\pm$ 0.12 & 2010.96$\pm$0.01 & 0.28$\pm$0.27 & 0.00 & 1.31 $\pm$ 0.24 \\
      & 22.0      & 35.79$\pm$0.16 & 2010.91$\pm$0.01 & 0.13$\pm$0.06 & -0.05$\pm$0.01 & \\
      & 14.5      & 12.63$\pm$0.06 & 2011.03$\pm$0.02 & 0.96$\pm$0.51 & 0.07$\pm$0.02 & \\
      & 8.0       & 14.27$\pm$0.24 & 2011.22$\pm$0.32 & 0.75$\pm$0.03 & 0.26$\pm$0.32 & \\
      & 4.8       & 7.80$\pm$0.28  & 2011.46$\pm$0.15 & 0.85$\pm$0.27 & 0.50$\pm$0.35 & \\ \hline
T     & 36.0      & 9.64$\pm$ 0.04 & 2013.78$\pm$0.01 & 0.50$\pm$0.49 & 0.00 & \\
      & 22.0      & 3.39$\pm$0.01 & 2013.81$\pm$0.12 & 0.30$\pm$0.92 & -0.03$\pm$0.12 & \\ \hline
\end{tabular}}
\caption{Gaussian fit based parameters, time lags and spectral index for segment 2 light curves.}
\label{tab2}
\end{table}

\begin{table}
\centering
\scalebox{.78}{
\begin{tabular}{llllllll} \hline 
Flare &  Frequency & $\Delta r$ & $\Omega_{r\nu }$ & $r_{core}$ & $B_{1pc}$ & $B_{core}$ \\
      & (GHz)      & ($\mu$as)     & (pc GHz$^{1/k_r}$)        & (pc)       & (G)       & (mG) \\ \hline
A & 4.8 & 90$\pm$16 & 3.97$\pm$0.74 & 34.84$\pm$0.83 & 0.37$\pm$0.05 & 11$\pm$2 \\
 & 8.0 & 80$\pm$20 & 6.63$\pm$1.67 & 34.47$\pm$1.82 & 0.55$\pm$0.10 & 16$\pm$4 \\ 
 & 14.5 & 26$\pm$19 & 5.16$\pm$3.72 & 14.54$\pm$4.02 & 0.45$\pm$0.25 & 31$\pm$23 \\
 & 22.0 & 53$\pm$10 & 23.76$\pm$4.52 & 43.62$\pm$4.90 & 1.43$\pm$0.20 & 33$\pm$6 \\ \hline
B & 4.8 & 42$\pm$17 & 1.85$\pm$0.76 & 16.26$\pm$0.83 & 0.21$\pm$0.06 & 13$\pm$5 \\ \hline
C & 4.8 & 66$\pm$8 & 2.91$\pm$0.39 & 25.55$\pm$0.45 & 0.30$\pm$0.03 & 12$\pm$2 \\
 & 8.0 & 26$\pm$17 & 2.18$\pm$1.41 & 11.32$\pm$1.52 & 0.24$\pm$0.12 & 21$\pm$14 \\
 & 14.5 & 14$\pm$16 & 2.81$\pm$3.13 & 7.93$\pm$3.38 & 0.29$\pm$0.24 & 36$\pm$41 \\
 & 22.0 & 14$\pm$2 & 6.48$\pm$0.91 & 11.90$\pm$0.98 & 0.54$\pm$0.06 & 45$\pm$6 \\ \hline
D & 4.8 & 6$\pm$16 & 0.26$\pm$0.71 & 2.32$\pm$0.76 & 0.05$\pm$0.10 & 21$\pm$56 \\
 & 8.0 & 42$\pm$26 & 3.47$\pm$2.15 & 18.00$\pm$2.33 & 0.34$\pm$0.16 & 19$\pm$12 \\
 & 14.5 & 37$\pm$22 & 7.27$\pm$4.30 & 20.49$\pm$4.66 & 0.59$\pm$0.26 & 29$\pm$17 \\
 & 22.0 & 30$\pm$2 & 13.50$\pm$0.93 & 24.79$\pm$1.02 & 0.94$\pm$0.05 & 38$\pm$3 \\ \hline
E & 8.0 & 53$\pm$25 & 4.36$\pm$2.07 & 22.63$\pm$2.24 & 0.40$\pm$0.14 & 18$\pm$9 \\
 & 14.5 & 44$\pm$20 & 8.68$\pm$3.91 & 24.46$\pm$4.24 & 0.67$\pm$0.23 & 27$\pm$13 \\
 & 22.0 & 19$\pm$11 & 8.64$\pm$4.95 & 15.86$\pm$5.36 & 0.67$\pm$0.28 & 42$\pm$25 \\ \hline
F & 22.0 & 22$\pm$6 & 9.72$\pm$2.71 & 17.85$\pm$2.93 & 0.73$\pm$0.02 & 41$\pm$ 12\\ \hline
G & 4.8 & 92$\pm$19 & 4.07$\pm$0.87 & 35.77$\pm$0.97 & 0.38$\pm$0.06 & 11$\pm$2 \\
 & 8.0 & 44$\pm$10 & 3.67$\pm$0.84 & 19.03$\pm$0.91 & 0.35$\pm$0.06 & 19$\pm$4 \\
 & 14.5 & 16$\pm$20 & 3.05$\pm$3.91 & 8.59$\pm$4.23 & 0.31$\pm$0.29 & 36$\pm$47 \\
 & 22.0 & 29$\pm$18 & 12.96$\pm$8.10 & 23.80$\pm$8.77 & 0.91$\pm$0.43 & 38$\pm$25 \\ \hline
H & 14.5 & 37$\pm$23 & 7.27$\pm$4.50 & 20.49$\pm$4.87 & 0.59$\pm$0.27 & 29$\pm$18 \\
 & 22.0 & 19$\pm$22 & 8.64$\pm$9.90 & 15.86$\pm$10.71 & 0.67$\pm$0.57 & 42$\pm$53 \\ \hline
I & 4.8 & 31$\pm$28 & 1.38$\pm$1.24 & 12.08$\pm$1.34 & 0.17$\pm$0.11 & 14$\pm$13 \\
 & 8.0 & 5$\pm$22 & 0.40$\pm$1.82 & 2.06$\pm$1.96 & 0.07$\pm$0.23 & 32$\pm$150 \\
 & 14.5 & 5$\pm$16 & 0.94$\pm$3.13 & 2.64$\pm$3.38 & 0.13$\pm$0.32 & 48$\pm$160 \\
 & 22.0 & 14$\pm$12 & 6.48$\pm$5.40 & 11.90$\pm$5.84 & 0.54$\pm$0.34 & 45$\pm$38 \\ \hline
J & 4.8 & 22$\pm$17 & 0.95$\pm$0.75 & 8.36$\pm$0.81 & 0.13$\pm$0.08 & 15$\pm$12 \\
 & 8.0 & 68$\pm$17 & 5.64$\pm$1.42 & 29.32$\pm$1.55 & 0.49$\pm$0.09 & 17$\pm$4 \\
 & 14.5 & 40$\pm$18 & 7.74$\pm$3.52 & 21.81$\pm$3.81 & 0.62$\pm$0.21 & 28$\pm$13 \\
 & 22.0 & 38$\pm$17 & 17.28$\pm$7.66 & 31.73$\pm$8.28 & 1.13$\pm$0.37 & 36$\pm$16 \\ \hline
K & 4.8 & 179$\pm$24 & 7.88$\pm$1.15 & 69.21$\pm$1.32 & 0.62$\pm$0.07 & 9$\pm$1 \\
 & 8.0 & 112$\pm$17 & 9.21$\pm$1.44 & 47.84$\pm$1.59 & 0.70$\pm$0.08 & 15$\pm$2 \\
 & 14.5 & 24$\pm$22 & 4.69$\pm$4.30 & 13.22$\pm$4.65 & 0.42$\pm$0.29 & 32$\pm$30 \\
 & 22.0 & 2$\pm$14 & 1.08$\pm$6.30 & 1.98$\pm$6.81 & 0.14$\pm$0.62 & 71$\pm$456 \\ \hline
L & 14.5 & 5$\pm$18 & 0.94$\pm$3.51 & 2.64$\pm$3.81 & 0.13$\pm$0.36 & 48$\pm$186 \\
 & 22.0 & 8$\pm$11 & 3.78$\pm$4.95 & 6.94$\pm$5.36 & 0.36$\pm$0.35 & 52$\pm$70 \\ \hline
M & 8.0 & 17$\pm$23 & 1.39$\pm$1.90 & 7.20$\pm$2.05 & 0.17$\pm$0.17 & 24$\pm$33 \\
 & 14.5 & 30$\pm$22 & 5.86$\pm$4.30 & 16.52$\pm$4.65 & 0.50$\pm$0.28 & 30$\pm$22 \\
 & 22.0 & 19$\pm$4 & 8.64$\pm$1.81 & 15.86$\pm$1.96 & 0.67$\pm$0.10 & 42$\pm$9 \\ 
 \hline
\end{tabular}}
\caption{Core position offsets, distance from jet base and magnetic field strengths inferred for segment 1 light curves.}   
\label{tab3}     
\end{table}

\begin{table}
\centering
\scalebox{.78}{
\begin{tabular}{llllllll} \hline
Flare &  Frequency & $\Delta r$ & $\Omega_{r\nu }$ & $r_{core}$ & $B_{1pc}$ & $B_{core}$ \\
      & (GHz)      & (mas)     & (pc GHz$^{1/k_r}$)         & (pc)       & (G)       & (G) \\ \hline
N & 4.8 & 53$\pm$10 & 1.80$\pm$0.36 & 22.11$\pm$0.87 & 0.21$\pm$0.03 & 9$\pm$2 \\
 & 8.0 & 41$\pm$10 & 2.40$\pm$0.60 & 19.44$\pm$1.40 & 0.26$\pm$0.05 & 13$\pm$3 \\ \hline
O & 14.5 & 62$\pm$6 & 8.01$\pm$0.80 & 39.77$\pm$1.90 & 0.63$\pm$0.05 & 16$\pm$2 \\
& 22.0 & 11$\pm$1 & 3.03$\pm$0.29 & 10.69$\pm$0.67 & 0.31$\pm$0.02 & 29$\pm$3 \\ \hline
P & 4.8 & 78$\pm$7 & 2.67$\pm$0.30 & 32.66$\pm$0.78 & 0.28$\pm$0.02 & 8$\pm$1 \\ \hline
Q & 4.8 & 76$\pm$19 & 2.59$\pm$0.67 & 31.66$\pm$1.60 & 0.27$\pm$0.05 & 9$\pm$2 \\ 
 & 8.0 & 26$\pm$5 & 1.56$\pm$0.30 & 12.58$\pm$0.71 & 0.19$\pm$0.03 & 15$\pm$3 \\
 & 14.5 & 19$\pm$10 & 2.47$\pm$1.29 & 12.24$\pm$2.98 & 0.26$\pm$0.10 & 21$\pm$11 \\
 & 22.0 & 28$\pm$8 & 7.73$\pm$2.25 & 27.31$\pm$5.22 & 0.62$\pm$0.13 & 23$\pm$7 \\ \hline
R & 14.5 & 38$\pm$2 & 4.93$\pm$0.29 & 24.47$\pm$0.71 & 0.44$\pm$0.02 & 18$\pm$1 \\
 & 22.0 & 29$\pm$2 & 8.07$\pm$0.59 & 28.50$\pm$1.38 & 0.64$\pm$0.03 & 22$\pm$2 \\ \hline
S & 4.8 & 60$\pm$38 & 2.05$\pm$1.31 & 25.13$\pm$3.04 & 0.23$\pm$0.11 & 9$\pm$6 \\
 & 8.0 & 31$\pm$2 & 1.84$\pm$0.14 & 14.86$\pm$0.36 & 0.21$\pm$0.01 & 14$\pm$1 \\
 & 14.5 & 8$\pm$1 & 1.08$\pm$0.13 & 5.35$\pm$0.31 & 0.14$\pm$0.01 & 26$\pm$3 \\
 & 22.0 & 6$\pm$1 & 1.68$\pm$0.28 & 5.94$\pm$0.66 & 0.20$\pm$0.02 & 33$\pm$6 \\ \hline
T & 22.0 & 4$\pm$1 & 1.01$\pm$0.28 & 3.56$\pm$0.65 & 0.13$\pm$0.03 & 38$\pm$10 \\
 \hline
\end{tabular}}
\caption{Core position offsets, distance from jet base and magnetic field strengths inferred for segment 1 light curves.}
\label{tab4}
\end{table}

\begin{figure}
\centerline{\includegraphics[scale=0.16]{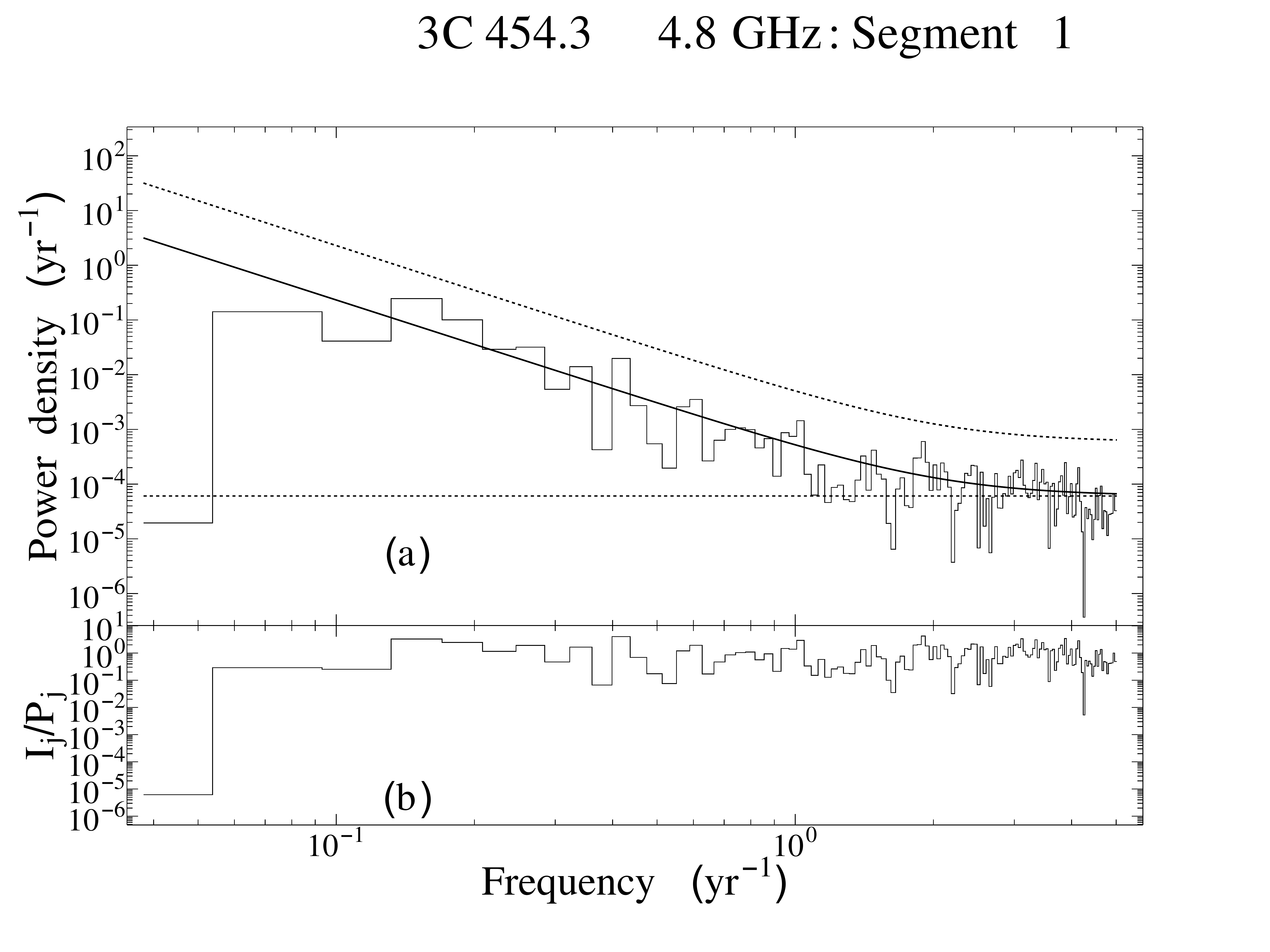}}
\centerline{\includegraphics[scale=0.16]{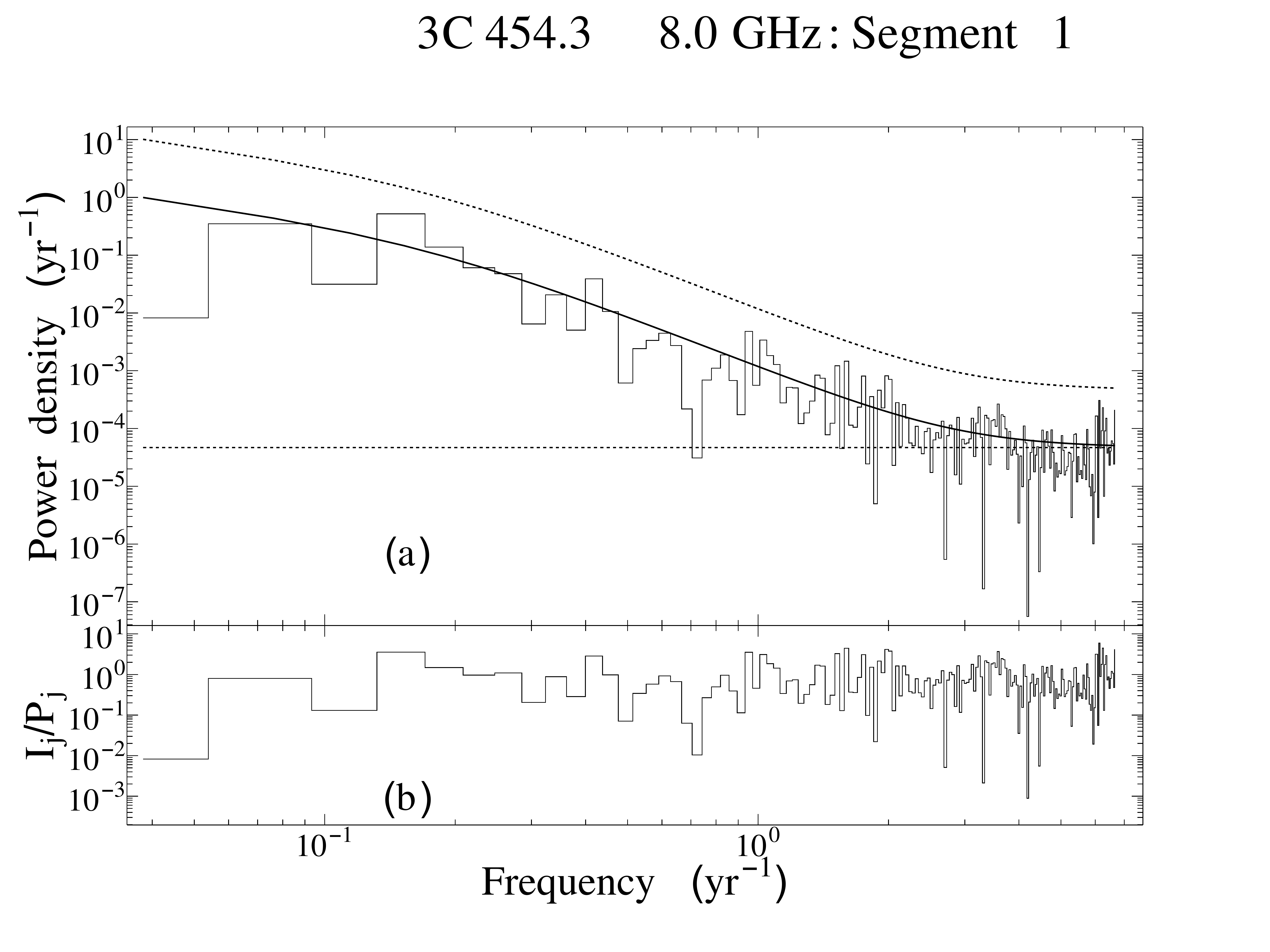}}
\centerline{\includegraphics[scale=0.16]{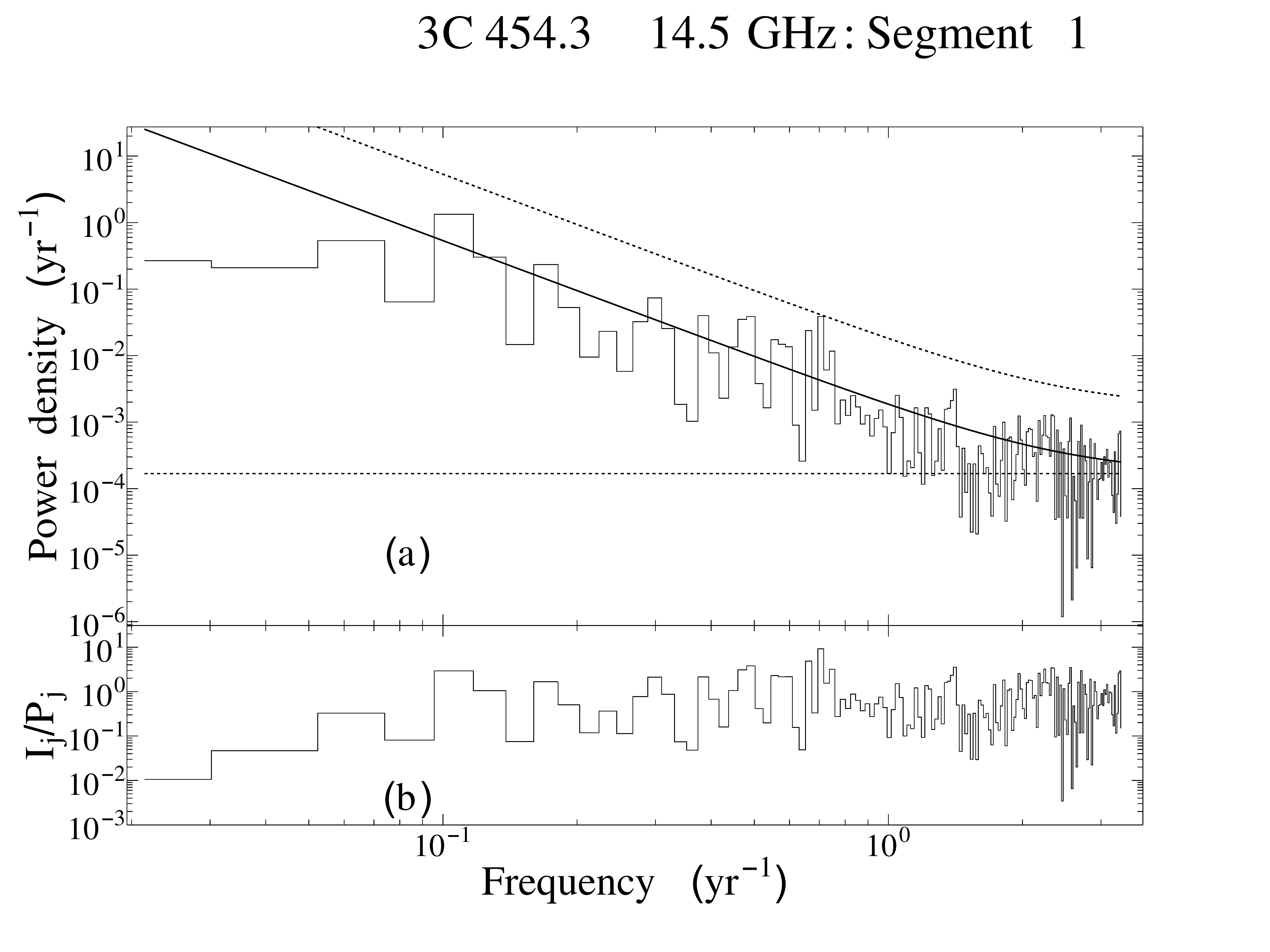}}
\caption{Periodogram analysis of segment 1 (beginning of observations to $\sim$ 2007.0) of the 4.8 GHz, 8.0 GHz and 14.5 GHz light curve flares. The best fit model is the solid curve, the dashed curve above it is the 99 \% significance contour which can identify statistically significant quasi-periodic components, the dot-dashed horizontal line is the white noise level and the plot below each periodogram panel shows the fit residuals.}
\label{seg1psda}
\end{figure}

\begin{figure}
\centerline{\includegraphics[scale=0.16]{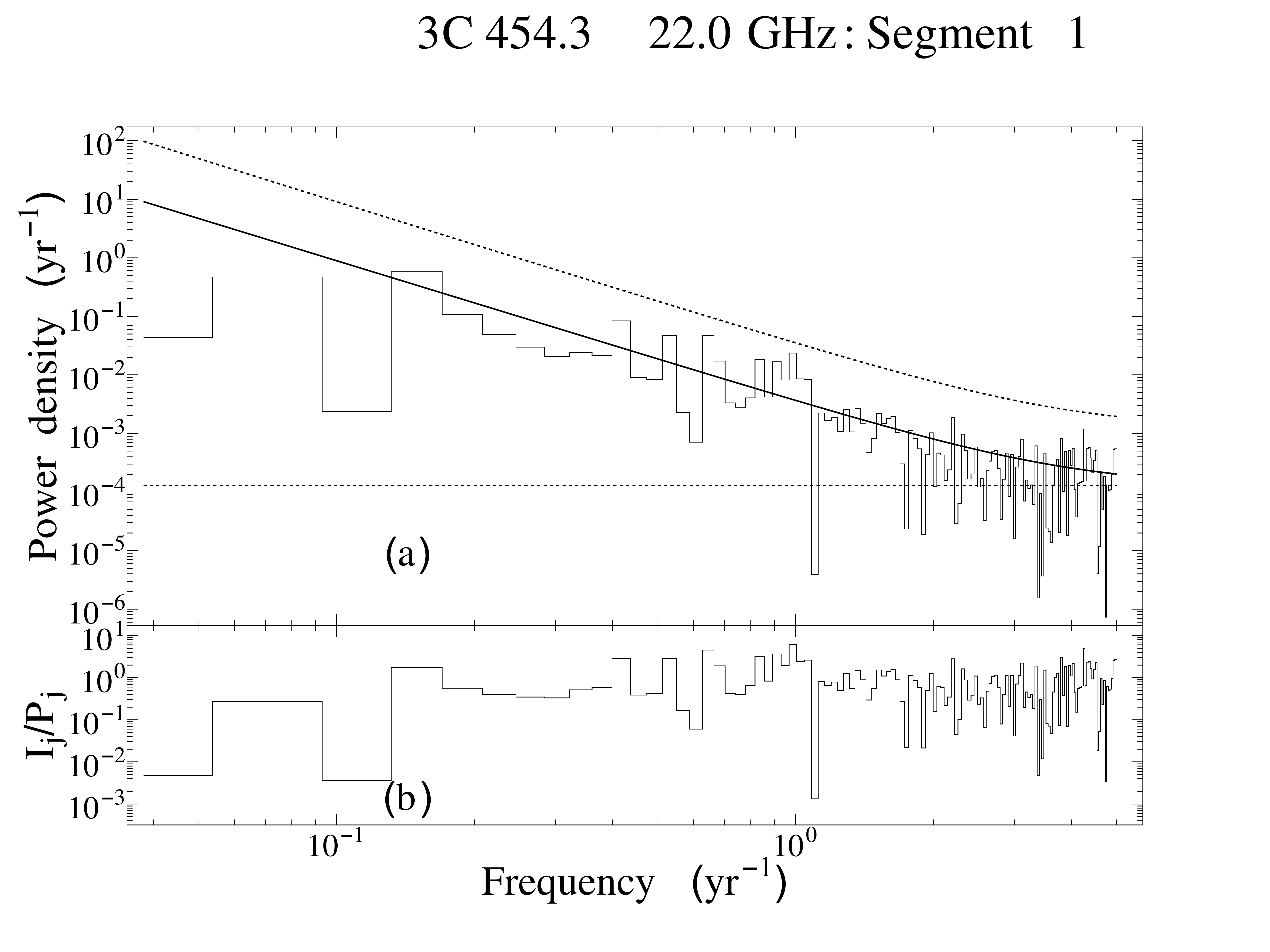}}
\centerline{\includegraphics[scale=0.16]{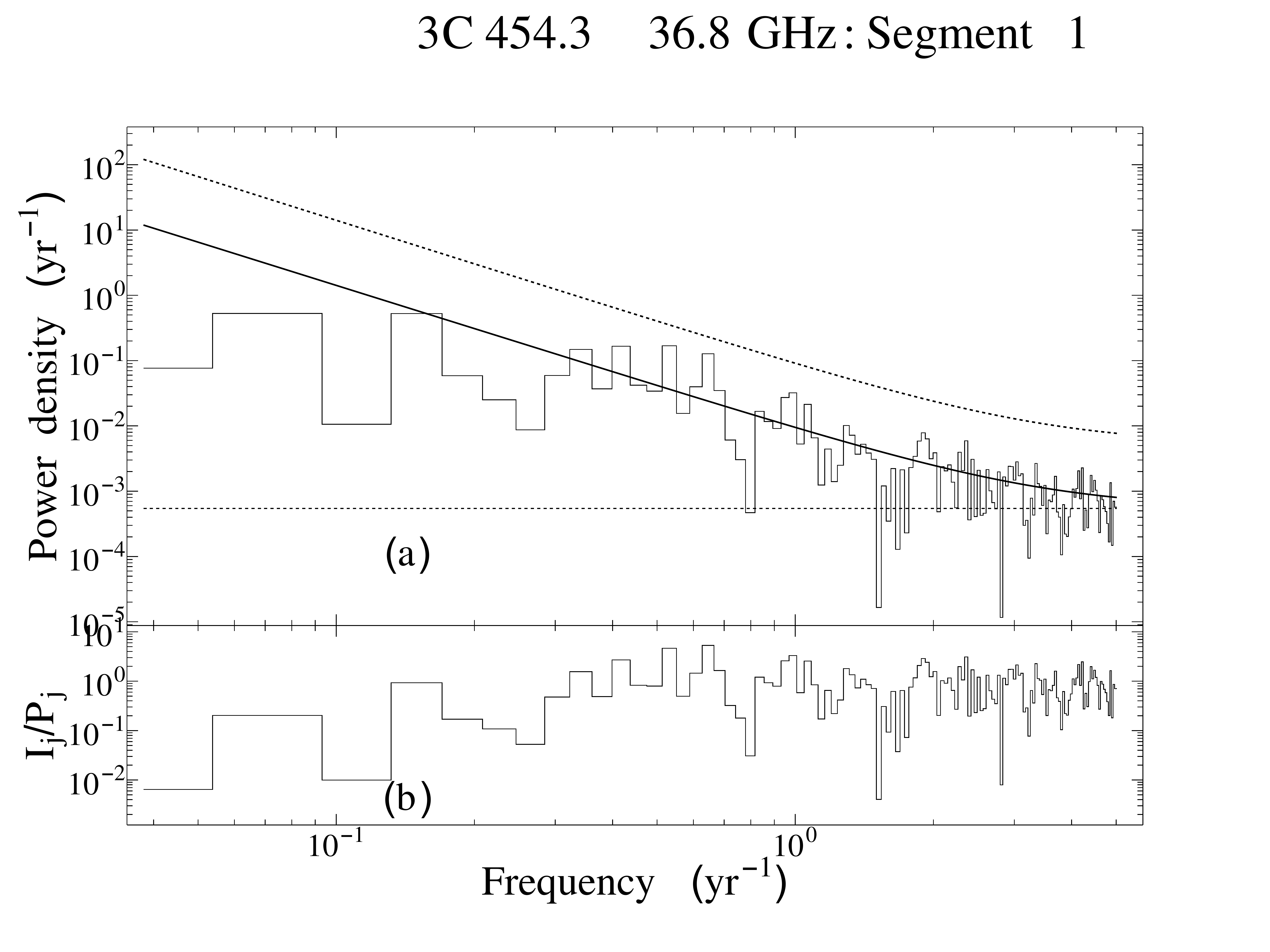}}
\caption{Periodogram analysis of segment 1 (beginning of observations to $\sim$ 2007.0) of the 22.0 GHz and 36.8 GHz light curve flares. The best fit model is the solid curve, the dashed curve above it is the 99 \% significance contour which can identify statistically significant quasi-periodic components, the dot-dashed horizontal line is the white noise level and the plot below each periodogram panel shows the fit residuals.}
\label{seg1psdb}
\end{figure}

\begin{figure}
\centerline{\includegraphics[scale=0.16]{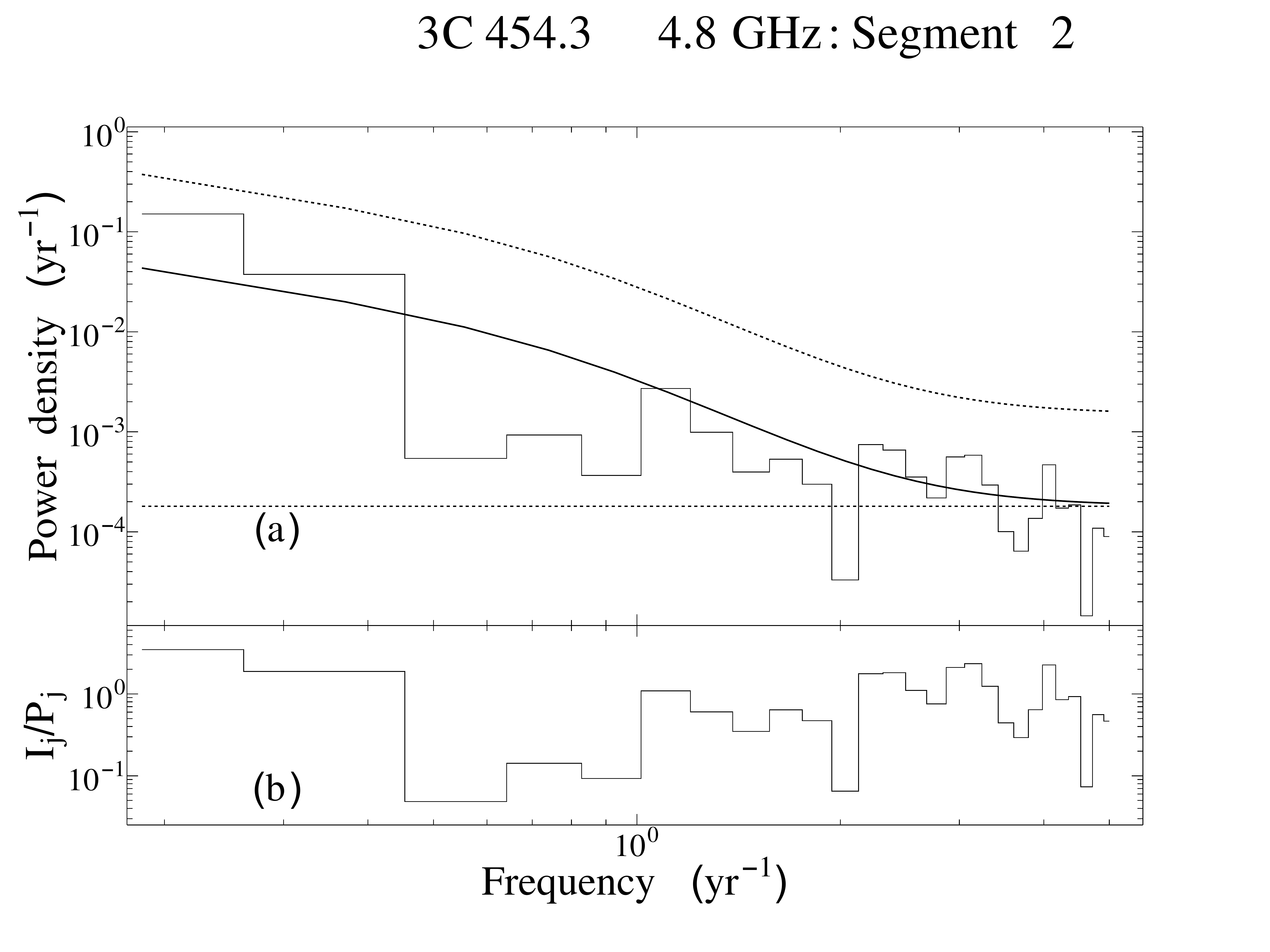}}
\centerline{\includegraphics[scale=0.16]{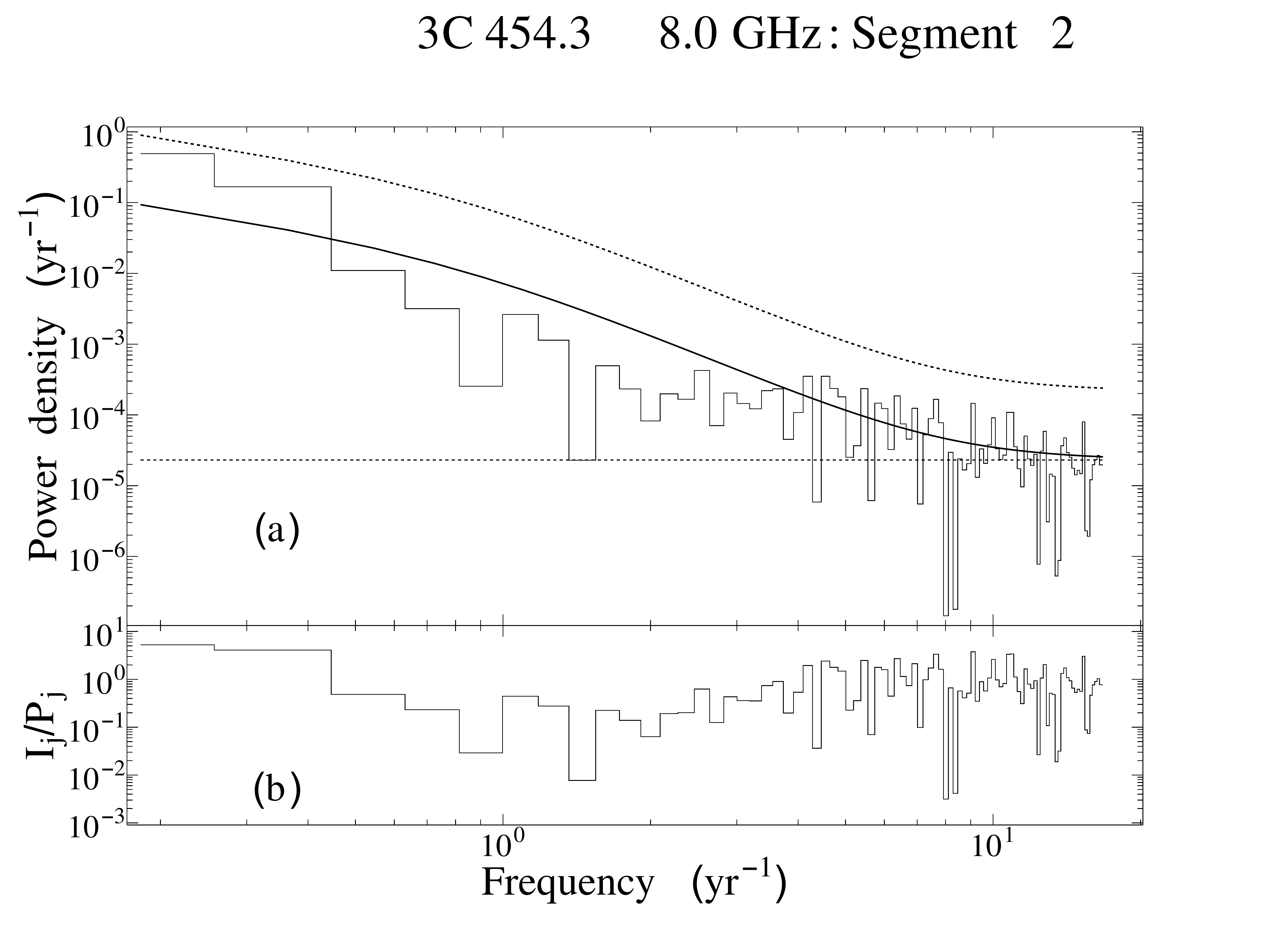}}
\centerline{\includegraphics[scale=0.16]{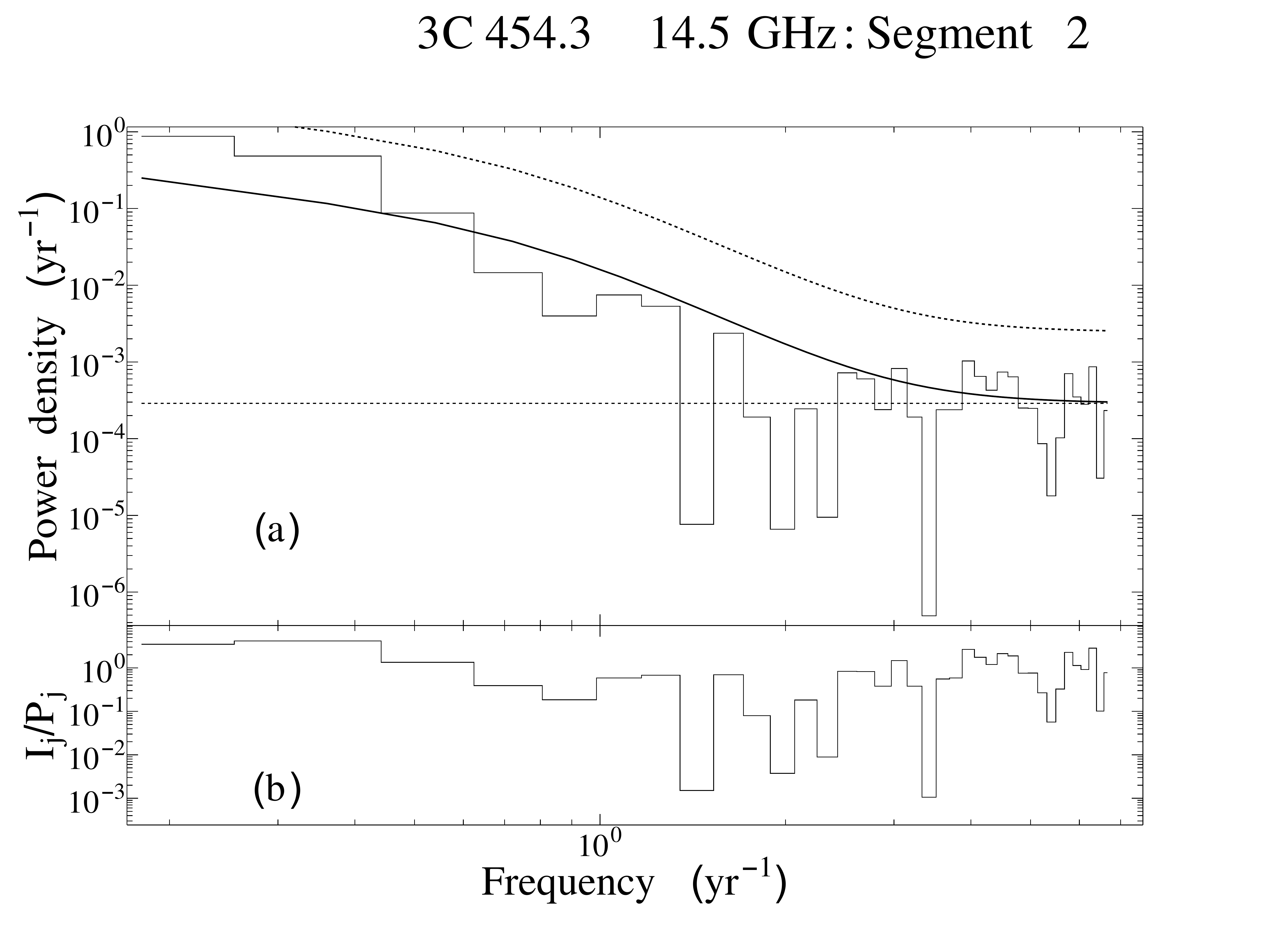}}
\caption{Periodogram analysis of segment 2 (2007.0 to end of observations) of the 4.8 GHz, 8.0 GHz and 14.5 GHz light curve flares. The best fit model is the solid curve, the dashed curve above it is the 99 \% significance contour which can identify statistically significant quasi-periodic components, the dot-dashed horizontal line is the white noise level and the plot below each periodogram panel shows the fit residuals.}
\label{seg2psda}
\end{figure}

\begin{figure}
\centerline{\includegraphics[scale=0.16]{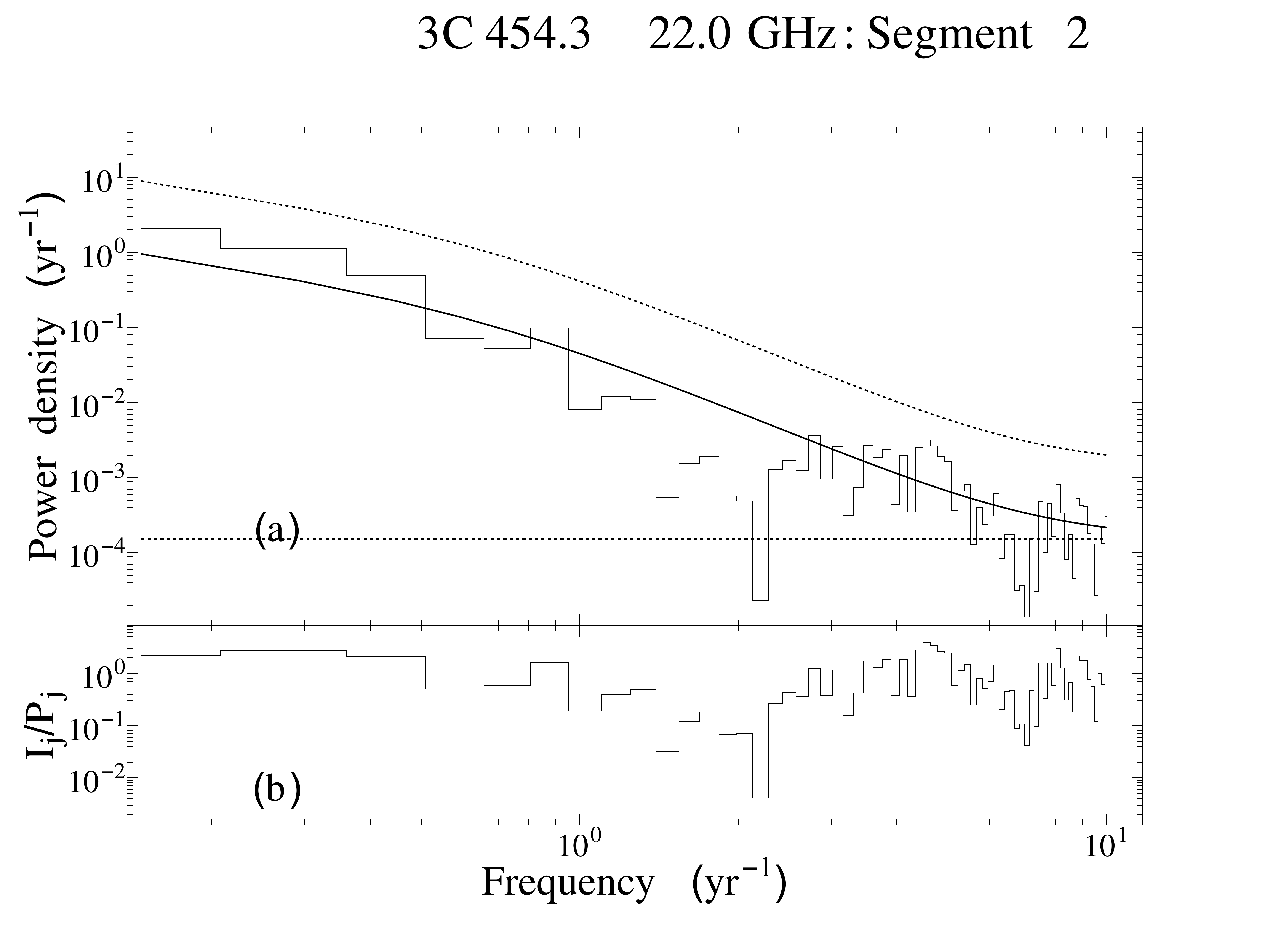}}
\centerline{\includegraphics[scale=0.16]{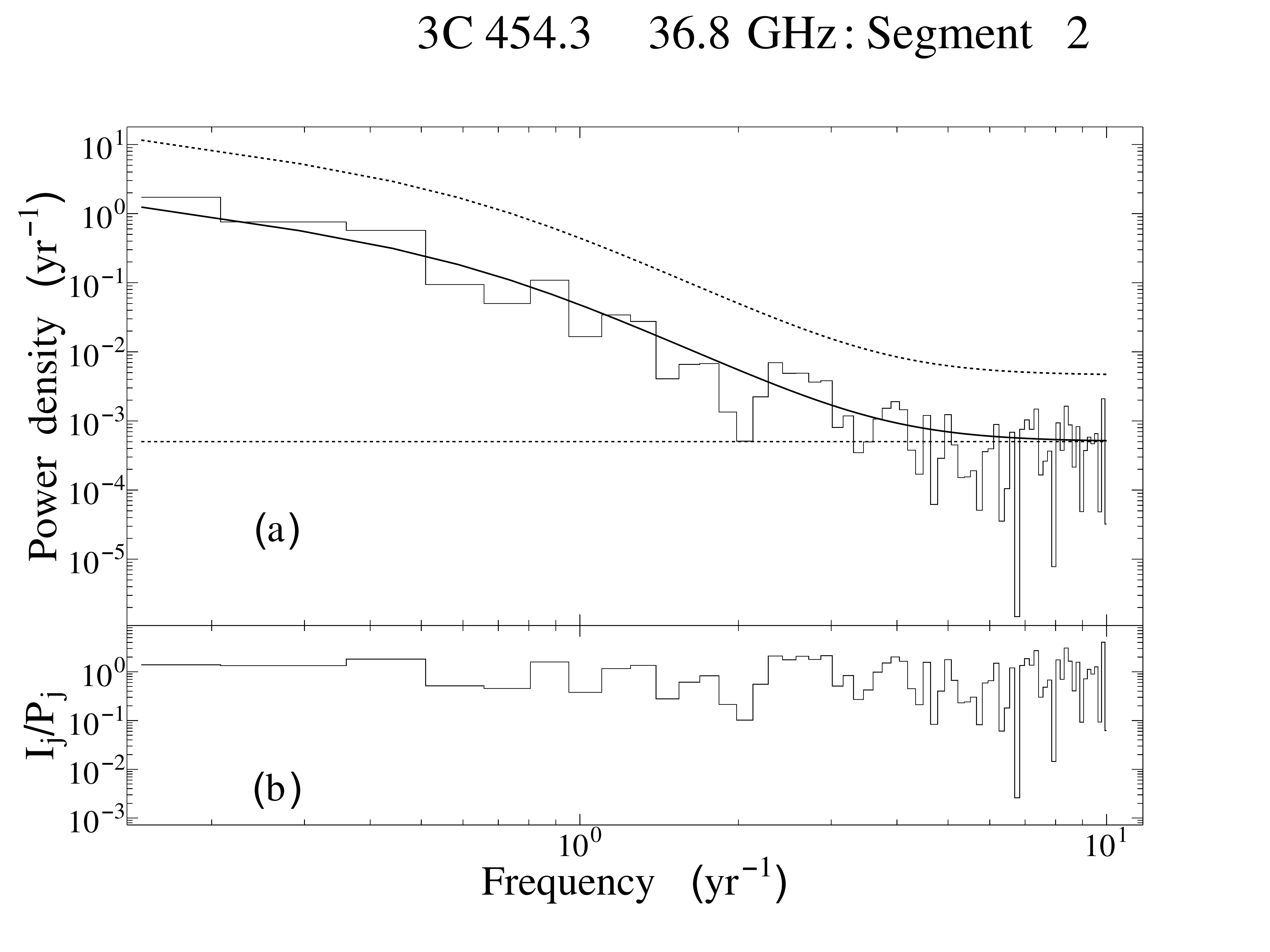}}
\caption{Periodogram analysis of segment 2 (2007.0 to end of observations) of the 22.0 GHz and 36.8 GHz light curve flares. The best fit model is the solid curve, the dashed curve above it is the 99 \% significance contour which can identify statistically significant quasi-periodic components, the dot-dashed horizontal line is the white noise level and the plot below each periodogram panel shows the fit residuals.}
\label{seg2psdb}
\end{figure}

In the time series analysis of these light curve segments using the periodogram, the power law model describes the PSD shape better in 4/10 segments. The power law slope ranges between $-1.6$ and $-3.0$ across all segments, consistent with that inferred for optical and X-ray light curve studies (e.g. \citealt{2012A&A...544A..80G,2014ApJ...791...74M}). There is no clear trend of increase or decrease of the slope with observation frequency indicating that the origin of the flares are from multiple propagating shocks as also indicated from the study on flare amplitude dependence on observation frequency. The bending power law model describes the PSD shape better in the remaining 6/10 segments. The power law slope at lower temporal frequencies ranges between $-2.5$ and $-3.5$. The slopes for a majority of the second segment light curves are consistently steep at $-3.5$; since the bending power law is the best fit for all second segment light curves, it indicates that the second segment flares originate from  regions of similar size. The timescale associated with the bend frequency ranges only between 0.51 yr and 0.66 yr for these second segment light curves, supporting the inference from the slopes with further information that the emitting region is at a similar distance from the base of the jet. The periodograms of each segment with the best fit PSD models are presented in Figs. \ref{seg1psda}, \ref{seg1psdb}, \ref{seg2psda} and \ref{seg2psdb} and detailed results from the time series analysis are presented in Table \ref{pgrambestfit}.

\begin{table}
\centering
\scalebox{.73}{
\begin{tabular}{lllllll}
\hline
Observation & PSD       & \multicolumn{3}{|l|}{PSD Fit parameters}    & AIC & Model \\
Frequency \&& model     & \multicolumn{3}{|l|}{} &     & likelihood\\
Segment     &           & log(A) & $\alpha$ & log(f$_b$)            &     & \\ \hline
4.8 GHz: Seg. 1 & {\bf PL} & -3.3 $\pm$ 0.1 & -2.7 $\pm$ 0.2 & & 1946.55 & 1.00 \\
 & BPL& -1.7 $\pm$ 0.2 & -2.9 $\pm$ 0.3 & -0.88 $\pm$ 0.21 & 1947.78 & 0.54 \\ 
4.8 GHz: Seg. 2 & PL & -2.8 $\pm$ 0.2 & -2.6 $\pm$ 0.4 & & 339.74 & 0.05 \\
 & {\bf BPL} & -2.0 $\pm$ 0.2 & -3.5 $\pm$ 0.4 & -0.19 $\pm$ 0.09 & 333.72 & 1.00 \\ \hline
8.0 GHz: Seg. 1 & PL & -2.9 $\pm$ 0.1 & -2.5 $\pm$ 0.2 & & 4301.24 & 0.17 \\
 & {\bf BPL} & -1.6 $\pm$ 0.1 & -2.5 $\pm$ 0.3 & -0.88 $\pm$ 0.16 & 4297.69 & 1.00 \\ 
8.0 GHz: Seg. 2 & PL & -2.6 $\pm$ 0.2 & -3.0 $\pm$ 0.3 & & 795.54 & $10^{-3}$ \\
 & {\bf BPL} & -1.7 $\pm$ 0.2 & -3.5 $\pm$ 0.2 & -0.20 $\pm$ 0.01 & 782.67 & 1.00 \\ \hline
14.5 GHz: Seg. 1 & {\bf PL} & -2.5 $\pm$ 0.1 & -2.3 $\pm$ 0.2 & & 1937.31 & 1.00 \\
 & BPL& -1.6 $\pm$ 0.1 & -3.5 $\pm$ 0.3 & -0.32 $\pm$ 0.12 & 1944.49 & 0.03 \\ 
14.5 GHz: Seg. 2 & PL & -2.1 $\pm$ 0.2 & -3.0 $\pm$ 0.3 & & 294.36 & 0.01 \\
 & {\bf BPL} & -1.3 $\pm$ 0.1 & -3.5 $\pm$ 0.2 & -0.20 $\pm$ 0.01 & 285.16 & 1.00 \\ \hline
22.0 GHz: Seg. 1 & {\bf PL} & -2.4 $\pm$ 0.1 & -2.4 $\pm$ 0.3 & & 1528.90 & 1.00 \\
 & BPL& -1.7 $\pm$ 0.2 & -3.5 $\pm$ 0.2 & -0.23 $\pm$ 0.12 & 1544.59 & $10^{-4}$ \\ 
22.0 GHz: Seg. 2 & PL & -1.6 $\pm$ 0.1 & -2.5 $\pm$ 0.3 & & 754.88 & 0.02 \\
 & {\bf BPL}& -0.8 $\pm$ 0.2 & -3.0 $\pm$ 0.3 & -0.18 $\pm$ 0.01 & 746.90 & 1.00 \\ \hline
36.8 GHz: Seg. 1 & {\bf PL} & -2.0 $\pm$ 0.1 & -1.9 $\pm$ 0.2 & & 1242.52 & 1.00 \\
 & BPL& -1.5 $\pm$ 0.2 & -3.3 $\pm$ 0.4 & -0.09 $\pm$ 0.17 & 1249.49 & 0.03 \\ 
36.8 GHz: Seg. 2 & PL & -1.5 $\pm$ 0.1 & -2.9 $\pm$ 0.3 & & 270.35 & 0.38 \\
 & {\bf BPL} & -0.8 $\pm$ 0.1 & -3.3 $\pm$ 0.3 & -0.29 $\pm$ 0.01 & 268.41 & 1.00 \\ \hline
\end{tabular}}
\caption{Results from the parametric PSD models fit to the periodogram. Columns 1 -- 7 give the observation frequency and segment, the model (PL: power law + constant noise, BPL: bending power law + constant noise), the best-fit parameters $\log(N)$, slope $\alpha$ and the bend frequency $f_b$ with their 95\% errors derived from $\Delta S$, the AIC and the likelihood of a particular model. The best fit PSD is highlighted.}
\label{pgrambestfit}
\end{table}

\section{Conclusions}
\label{conclusions}

The core shift effect in the parsec scale jet of the blazar 3C 454.3 was studied in the 4.8 GHz - 36.8 GHz radio wavelengths. The inferred time delays $\Delta t$ for flares in these frequencies are in agreement with the frequency dependence of synchrotron emission from the core (surface with optical depth $\tau = 1$) (\citealt{1981ApJ...243..700K}). From the fit $\Delta t \propto \nu^{-1/k_r}$ where $\nu$ are the observation frequencies, we obtain a weighted mean $k_r = 1.10 \pm 0.18$, consistent with the equipartition between the magnetic field energy density and the particle energy density. The large scale magnetic field thus dominates the kinematics of parsec scale jets as its decay is slower compared to the decay in the particle number density as $B_{\rm core} \propto r^{-1}$ and $N \propto r^{-2}$. From the fit to the flare amplitudes $A\propto \nu^{\alpha}$, we infer spectral indices $\alpha$ in the range $-0.24$ and $1.52$ for both segments. As the shock propagates downstream, it loses energy through interactions with the plasma. Thus, the amplitude of flares at lower frequencies is expected to be reduced. The opposite trend, i.e., decreasing amplitude with increasing frequency, can occur if the core is excited by  shocks that have sufficient energy to cause flaring at lower frequencies but are not energetic enough to cause excitation upstream at higher frequencies. In our study, we obtain both positive and negative $\alpha$. The flaring activity can thus be attributed to multiple propagating shocks.

We infer a weighted mean $B_1 = 0.5 \pm 0.2$ G and $B_{\rm core} = 46 \pm 16$ mG considering both segments. These results are consistent within error bars of $B_1 = 0.493$ G obtained in \cite{2012ApJ...759..114C} and $B_{\rm core} = 0.04 \pm 0.02$ G obtained in \cite{2014MNRAS.437.3396K}. We obtain a weighted mean $\Omega_{r\nu} = 6.4 \pm 2.8$ pc GHz$^{1/k_r}$ when averaged over all frequency pairs in both segments. The source was also studied by \cite{2012A&A...545A.113P} where estimates of $\Omega_{r\nu}$, $B_1$ and $B_{\rm core}$ for the 15/8 GHz frequency pair are 22 pcGHz ($k_r = 1$ was assumed), 1.13 G and 0.06 G respectively. The inferred $B_1$ and $B_{\rm core}$ range in our study are consistent with these earlier estimates. The shift $\mu = 0.70$ mas/yr was assumed in \cite{2014MNRAS.437.3396K} and as a result, $\Omega_{r\nu} = 43 \pm 10$ pc GHz was obtained. As it was noted that the chosen value is $2-8$ times higher than that obtained in previous studies, we used a lower limit of $\mu = 0.12$ mas/yr of a range 0.12 - 0.53 mas/yr \citep{2005AJ....130.1418J} at 43 GHz, and 0.3 mas/yr at 15 GHz \citep{2013AJ....146..120L} to estimate $\Omega_{r\nu}$ which is the reason for our estimate being $\sim 4 - 15$ times smaller than that obtained previously. Based on the statistical trend shown by the $r_{\rm core}$ as a function of $\nu$, we infer that the synchrotron opacity model may not be valid for all cases.


From the time series analysis we obtain typical power law slopes in the range $-1.6$ to $-3.5$ for the power law as well as bending power law PSD shapes at lower temporal frequencies. The analysis indicates that the first segment light curves are consistent with multiple shock excitation events. The bending power law model is inferred to be a better PSD fit with a $-3.5$ slope and consistent bend timescales ranging between 0.51 yr and 0.66 yr for the second segment light curves indicating that the emitting cores for these flares are similar in size and distance from the jet base.

Important objectives met in this study include the demonstration of the computational efficiency and statistical basis of the analysis technique involving the piecewise Gaussian fit, which is automated in the determination of relevant flares; the ability to use this analysis across a wide range of light curves; the check for and confirmation of consistency of the results obtained for both light curve segments; the check for consistency with previously reported results based on a single flare in \cite{2014MNRAS.437.3396K} and the evidence it provides for the core shift dependence on the frequency and its usefullness in the determination of a variety of jet diagnostics in the region close to the resolving limit of VLBI (in the near vicinity of the central supermassive black hole).

\section{Acknowledgements}

We thank the referee for a careful reading of the manuscript and suggesting changes, additions and clarifications where necessary, thus improving our paper. The work at ARIES is partially supported by India-Ukraine inter-governmental project ``Multiwavelength Observations of Blazars", No. INT/UKR/2012/P-02 funded by the Department of Science and Technology (DST), Government of India. A.C.G. work is partially supported by the Chinese Academy of Sciences Visiting Fellowship for Researchers from Developing Countries (grant no. 2014FFJA0004). UMRAO was funded by a series of grants from the US National Science Foundation and from the NASA Fermi GI program. M.F.G. acknowledges the supports from the National Science Foundation of China (grant 11473054) and the Science and Technology Commission of Shanghai Municipality (14ZR1447100). Figures \ref{seg1ga}, \ref{seg1gb}, \ref{seg2ga}, \ref{seg2gb}, \ref{seg1psda}, \ref{seg1psdb}, \ref{seg2psda} and \ref{seg2psdb} in this manuscript were created using the LevelScheme scientific figure preparation system \citep{2005CoPhC.171..107C}.

\bibliography{main}

\end{document}